\documentclass[useAMS,usenatbib, proof]{mn2e}

\usepackage{graphicx}
\usepackage{times}
\usepackage{amssymb,amsmath}
\usepackage[draft]{hyperref}
\usepackage{wasysym}
\usepackage{textcomp}

\newcommand{\beq}{\begin{equation}}
\newcommand{\eeq}{\end{equation}}

\def\hii{\hbox{H~\textsc{ii}}}

\newcommand{\lsim}{\ \raise
-2.truept\hbox{\rlap{\hbox{$\sim$}}\raise5.truept\hbox{$<$}\ }}
\newcommand{\gsim}{\ \raise
-2.truept\hbox{\rlap{\hbox{$\sim$}}\raise5.truept\hbox{$>$}\ }}
\newcommand{\simsim}{\ \raise
-2.truept\hbox{\rlap{\hbox{$\sim$}}\raise5.truept\hbox{$\sim$}\ }}

\def\gtorder{\mathrel{\raise.3ex\hbox{$>$}\mkern-14mu
                \lower0.6ex\hbox{$\sim$}}}
\def\ltorder{\mathrel{\raise.3ex\hbox{$<$}\mkern-14mu
                \lower0.6ex\hbox{$\sim$}}}

\def\arcsec{\hbox{$^{\prime\prime}$}}
\def\solar{\mbox{$_{\normalsize\odot}$}}

\def\deg{\hbox{$^\circ$}}

\newcommand{\degdot}{\hbox{\rlap{\hbox{$.$}}\hbox{$\deg$}}}

\def\aj{AJ}                   
             
\def\apj{ApJ}                 
\def\apjl{ApJL}                
\def\apjs{ApJS}               
           
\def\apss{Ap\&SS}             
\def\aap{A\&A}

\def\fcp{Fundam. Cosm. Phys.}

\def\mnras{MNRAS}

\def\pasp{PASP}

\def\fcp{Fund.~Cosmic~Phys.}

\def\physscr{Phys.~Scr.}

\title{Hierarchical Star Formation across the  ring galaxy NGC\,6503} 

\author[D. A. Gouliermis et al.]
  {Dimitrios A. Gouliermis,$^{1,2,}$\thanks{gouliermis@uni-heidelberg.de}
   David Thilker,$^{3}$ 
   Bruce G. Elmegreen,$^{4}$
   Debra M. Elmegreen,$^{5}$
 \newauthor
   Daniela Calzetti,$^{6}$
   Janice C. Lee,$^{7,8}$
   Angela Adamo,$^{9}$
   Alessandra Aloisi,$^{7}$ 
   Michele Cignoni,$^{7}$
 \newauthor
   David O. Cook,$^{10}$
   Daniel Dale,$^{10}$
   John S. Gallagher III,$^{11}$
   Kathryn Grasha,$^{6}$   
   Eva K. Grebel,$^{12}$
 \newauthor
   Artemio Herrero Dav\'{o},$^{13}$ 
   Deidre A. Hunter,$^{14}$
   Kelsey E. Johnson,$^{15}$
   Hwihyun Kim,$^{16,17}$
 \newauthor
   Preethi Nair,$^{18}$
   Antonella Nota,$^{7}$ 
   Anne Pellerin,$^{19}$
   Jenna Ryon,$^{11}$
   Elena Sabbi,$^{7}$ 
 \newauthor
    Elena Sacchi,$^{20,21}$
   Linda\,J. Smith,$^{22}$
   Monica Tosi,$^{21}$
   Leonardo Ubeda,$^{7}$
   Brad Whitmore$^{7}$ \\ 
  \\
      $~~^1$University of Heidelberg, Centre for Astronomy, Institute for Theoretical Astrophysics, Albert-Ueberle-Str.\,2, 69120 Heidelberg, Germany\\
      $~~^2$Max Planck Institute for Astronomy,  K\"{o}nigstuhl\,17, 69117 Heidelberg, Germany \\
      $~~^3$Department of Physics and Astronomy, Johns Hopkins University, 3701 San Martin Drive, Baltimore, MD 21218, USA\\
      $~~^4$IBM Research Division, T.J. Watson Research Center, Yorktown Hts., NY 10598, USA \\
      $~~^5$Vassar College, Dept. of Physics and Astronomy, Poughkeepsie, NY 12604, USA \\
      $~~^6$Department of Astronomy, University of Massachusetts -- Amherst, Amherst, MA 01003, USA \\
      $~~^7$Space Telescope Science Institute, 3700 San Martin Drive, Baltimore, MD 21218, USA \\
      $~~^8$Visiting Astronomer, Spitzer Science Center, Caltech. Pasadena, CA, USA \\
      $~~^9$Department of Astronomy, Oskar Klein Centre, Stockholm University, AlbaNova University Centre, SE-106 91 Stockholm, Sweden \\
  $^{10}$Department of Physics and Astronomy, University of Wyoming, Laramie, WY, USA \\
  $^{11}$Department of Astronomy, University of Wisconsin-Madison, WI 53706, USA \\
  $^{12}$Astronomisches Rechen-Institut, Zentrum f\"ur Astronomie der Universit\"at Heidelberg, M\"onchhofstr. 12-14, D-69120 Heidelberg, Germany \\
  $^{13}$Instituto de Astrof'sica de Canarias, Research Division, C/ V\'{i}a L\'{a}ctea s/n, 38200 La Laguna, Tenerife, Spain \\
  $^{14}$Lowell Observatory, 1400 West Mars Hill Road, Flagstaff, AZ 86001, USA \\
  $^{15}$Department of Astronomy, University of Virginia, P.O. Box 400325, Charlottesville, VA 22904-4325, USA \\
  $^{16}$School of Earth and Space Exploration, Arizona State University, Tempe, AZ 85287, USA \\
  $^{17}$Korea Astronomy and Space Science Institute, Daejeon, Republic of Korea \\
  $^{18}$Department of Physics and Astronomy, University of Alabama, Tuscaloosa, AL, USA \\
  $^{19}$Department of Physics and Astronomy, State University of New York at Geneseo, Geneseo, NY, USA \\
  $^{20}$Department of Physics and Astronomy, Bologna University, Bologna, Italy \\
  $^{21}$INAF-Osservatorio Astronomico di Bologna, Via Ranzani 1, I-40127 Bologna, Italy \\
  $^{22}$European Space Agency and Space Telescope Science Institute, 3700 San Martin Drive, Baltimore, MD 21218, USA 
  }

\begin{document}

\date{Draft version \today}

\pagerange{\pageref{firstpage}--\pageref{lastpage}} \pubyear{2015}

\maketitle

\label{firstpage}


\begin{abstract}
We present a detailed clustering analysis of the young stellar population across the star-forming ring galaxy NGC\,6503, 
based on the deep HST photometry obtained with the {\em Legacy ExtraGalactic UV Survey} (LEGUS). 
We apply a contour-based map analysis technique and identify in the stellar surface density map 244 distinct star-forming structures at various levels of significance. 
These stellar complexes are found to be organized in a hierarchical fashion with 95\% being 
members of three dominant super-structures located along the star-forming ring. The size distribution of the identified structures and the 
correlation between their radii and numbers of stellar members show power-law behaviors, as expected from scale-free processes. The self-similar distribution of
young stars is further quantified from their autocorrelation function, with a fractal dimension 
of $\sim$\,1.7 for length-scales between $\sim$\,20\,pc and 2.5\,kpc. The young stellar radial distribution sets the extent of the star-forming 
ring at radial distances between 1 and 2.5\,kpc. About 60\% of the young stars belong to the detected stellar structures, while the remaining 
stars are distributed among the complexes, still inside the ring of the galaxy.  The analysis of the time-dependent clustering of young populations 
shows a significant change from a more clustered to a more distributed behavior in a time-scale of $\sim$\,60\,Myr. 
The observed hierarchy in stellar clustering is consistent with star formation being regulated by turbulence across the ring. The rotational velocity difference between the 
edges of the ring suggests shear as the driving mechanism for this process. Our findings reveal the interesting case of an inner ring forming stars in a hierarchical fashion.
\end{abstract}




\begin{figure*}
\centering
\includegraphics[width=1.\textwidth]{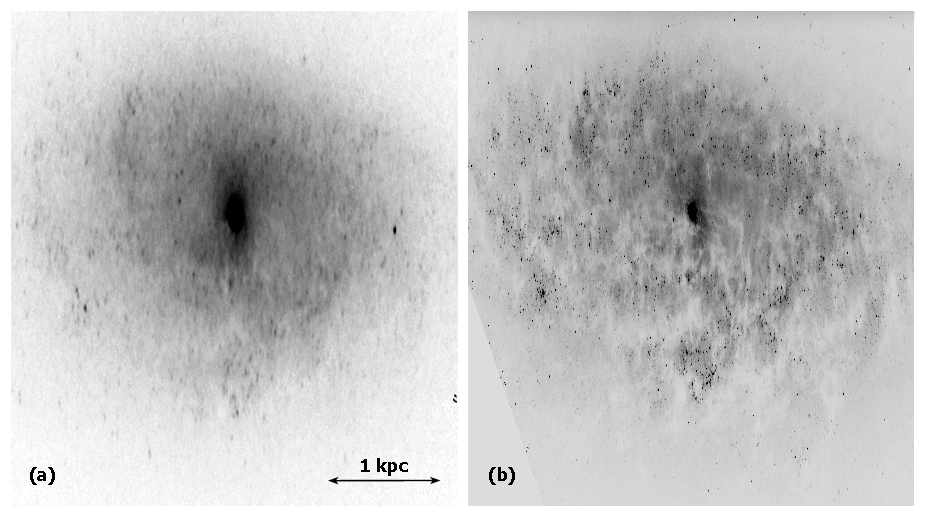} 
\caption{De-projected images of NGC\,6503. (a) Ground-based H-band (1.6 $\mu$m) image obtained  by members of our team 
with the WHIRC NIR imager at the WIYN 3.5-m telescope. (b) HST LEGUS image in the WFC3 filter F555W (530.8 nm).
These images provide evidence for the bar of the galaxy and indicate the overall presence of spiral structure with its weak arms. { Details on the deprojection technique are given in Sect.\,\ref{s:deproject}}.
 \label{fig:ngc6503ima}}
\end{figure*}

\section{Introduction}

Star formation, the conversion of gas into stars, is a key  process in shaping the structure, morphology and evolution of galaxies.
Young star clusters, OB associations and large complexes of young stars are the signposts of the recent star formation across a galaxy.
These various young stellar concentrations, covering a large dynamic range in physical length-scales, do not appear to be independent from 
each other but rather related to each other in a hierarchical fashion. Large loose structures of stars host smaller and more 
compact star-forming systems, which themselves are sub-structured \citep[e.g.,][]{elmegreen-ppiv}. The characterization of this clustering behavior 
and its origins are key issues in understanding how star formation progresses in space and time across galactic scales. Samples of resolved young stellar populations 
across a whole galaxy allow the detailed investigation of their formation in structures, improving our understanding from integrated stellar 
light. 

There are only a few investigations on the clustering of resolved young stars across galactic scales. They 
are focused on the Magellanic Clouds \citep[][]{maragoudaki98, gieles08, bastian09lmc}, NGC\,6822 \citep{gouliermis10},
and M\,33 \citep{bastian07}, up to sub-Mpc distances, where individual stars could be resolved from the ground. These studies give 
evidence of the self-similar scaling relations in the parameters of the identified stellar structures, demonstrating the hierarchical 
nature of stellar structural morphology on galactic scales. With the present study we extend these investigations to 
larger distances in the Local Volume, based on the exquisite resolving ability of the {\sl Hubble Space Telescope} 
(HST) with its cameras {\sl Advanced Camera for Surveys} (ACS) and {\sl Wide-Field Camera 3} (WFC3). We also
add to the sample of investigated galaxies the interesting case of the ring (possibly barred) spiral galaxy NGC\,6503 
(Fig.\,\ref{fig:ngc6503ima}). 

{ NGC\,6503  is classified as of morphological type S\underline{A}B(s)bc:\footnote{This type is according to the  classification by \cite{buta15} from the {\em Spitzer Survey of Stellar Structure in Galaxies} (S$^{4}$G). The galaxy was previously classified according to the {\em Third Reference Catalog of Bright Galaxies} \citep{devaucouleurs91} as SA(s)cd, i.e., a non-barred spiral with loosely wound spiral arms, and with no apparent ring.}, i.e., a pure s-shaped spiral galaxy, with possibly well-developed spiral arms, showing a trace of a bar \citep{buta15}.} The galaxy has also a patchy circumnuclear appearance in H$\alpha$, interpreted by \cite{knapen06} as a nuclear star-forming ring.
The same authors postulate that while this galaxy is possibly one of the rare cases of ring spirals classed as 
unbarred\footnote{Rings are most probably resonance phenomena, caused by a rotating bar or other non-axisymmetric 
disturbance in the disk. The current evidence supports the idea that ``rings are a {\em natural} consequence of 
barred galaxy dynamics'' \citep{butacombes96}.}, it is most likely to actually be barred. Indeed, both the H{\sc i} 
velocity field \citep{bottema97} and near-IR imaging \citep{freeland10} support the presence of a strong end-on bar. 
Moreover, \cite{freeland10} argue that the previously identified nuclear ring is instead an inner ring spanned in diameter 
by the bar (Fig.\,\ref{fig:ngc6503ima}). According to theory, rings form by gas accumulation through the action of gravitational 
torques from the bar pattern \citep[e.g.,][]{simkin80, schwarz84}. They are thus sites of active star formation in the 
galaxy \citep[][]{butacombes96}. 

\begin{figure*}
\centering
\includegraphics[width=\textwidth]{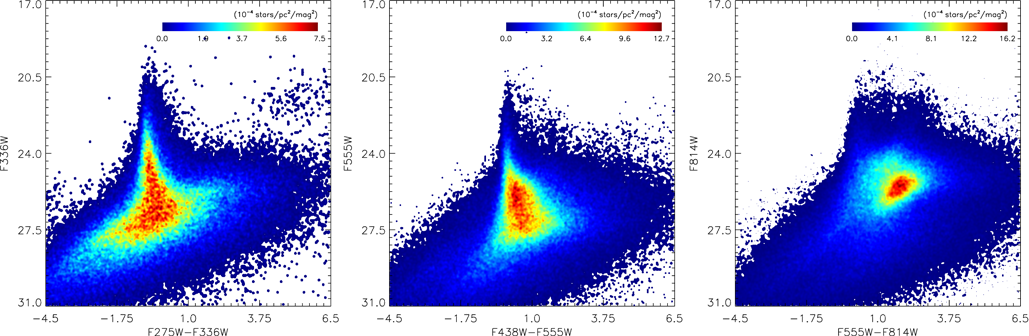} 
\caption{Hess diagrams of the observed stellar populations across the extent of NGC\,6503. 
Three diagrams are shown for stars identified in three different LEGUS filter-pairs with no quality 
cuts. Stellar surface densities across the diagrams are indicated by the color bars. 
\label{fig:hess_diagrams}}
\end{figure*}


Being a ring galaxy observed by HST, NGC\,6503 is a unique case for investigating  in detail star formation across an inner ring. 
Specific questions that can be probed are: (a) What is the length-scale of stellar structures that can be 
formed in the ring, and for how long can they survive? (b) Is star formation hierarchical across the ring? (c) What is
the role of ring dynamics in the star formation process?  In order to answer these questions we use the rich census of young blue stars resolved 
with HST in NGC\,6503, and we investigate their 
recent star formation as imprinted in their clustering morphology. Our goal is to quantify the young stellar spatial distribution and characterize 
its clustering on galactic scales, identify any characteristic scale of star formation or confirm its scale-free nature, determine the time-scale for large-scale
structure survival, and resolve its connection to the dynamics of the galaxy. 

Located at a distance of about 5\,Mpc \citep{karachentsev03}, NGC\,6503 is far enough for its entire extent to be mostly covered in a 
single WFC3 field-of-view, but also close enough for its young stellar content (down to $\sim$\,5\,M$_\odot$) to be sufficiently resolved 
with HST. The galaxy is also located away from the Galactic plane ($b=$\,30\degdot64) and therefore does not suffer from 
significant foreground extinction from the Milky Way, which corresponds to a reddening of $E(B-V) \simeq$\,0.03\,mag 
\citep{schlaflyfinkbeiner11}.  With this analysis we demonstrate the great advancement in resolved extragalactic stellar populations studies, accomplished with the 
{\em Legacy ExtraGalactic UV Survey}\footnote{\href{https://legus.stsci.edu/}{https://legus.stsci.edu/}} 
(LEGUS), aimed at the investigation of star formation and its relation with galactic environment in 50 nearby star-forming galaxies \citep[][]{calzetti15}.

This paper is organized as the following. In Sect.\,\ref{s:obs} we describe the LEGUS dataset of NGC\,6503 
and its photometry. We also select the stellar samples corresponding to the young blue populations of the galaxy,
and address their spatial distribution in comparison to that of the more evolved populations. In Sect.\,\ref{s:structident} 
we perform the identification of large young stellar structures across the galaxy, and present their structure tree to
visualize their hierarchical morphology. In this section we also discuss the demographics of the structural parameters 
of the identified systems and investigate 
their parameters correlations in an attempt to identify any characteristic scale of star formation or the power-law behavior
expected from scale-free processes. In Sect.\,\ref{s:hierarchy} we study the global spatial distribution of the stars through their 
surface density profiles and quantify their hierarchical distribution across the whole galaxy with the construction 
of their two-point (auto-)correlation function. In the same section we determine the time-scale
for the evolution of stellar structures through the study of the pair separations and the minimum spanning trees 
of stars as a function of their indicative evolutionary ages. In Sect.\,\ref{s:disc} we discuss our findings in terms
of galactic dynamics across the star-forming ring. We summarize our results in Sect.\,\ref{s:sum}.

\section{Observations and Photometry}\label{s:obs}

LEGUS is a HST panchromatic stellar survey of 50 nearby star-forming dwarf and spiral galaxies 
with an emphasis on UV-enabled science applications. Images in a 
wide waveband coverage from the near-UV to the I-band are being collected with WFC3 and ACS in parallel, 
and combined with archival optical ACS data.
The survey, its scientific objectives and the data reduction are described in \cite{calzetti15}.
Stellar photometry will be described in detail in Sabbi et al. (in prep). 
The images of NGC\,6503 to be used in our analysis were obtained in the filters F275W, 
F336W, F438W, F555W and F814W (equivalent to NUV, U, B, V, and I respectively) in August 2013. 
We applied the pixel-based correction for charge-transfer efficiency (CTE) degradation using tools 
provided by STScI, before processing them with {\sc astrodrizzle} and prior to their photometry.

\begin{figure}
\centering
\includegraphics[width=0.85\columnwidth]{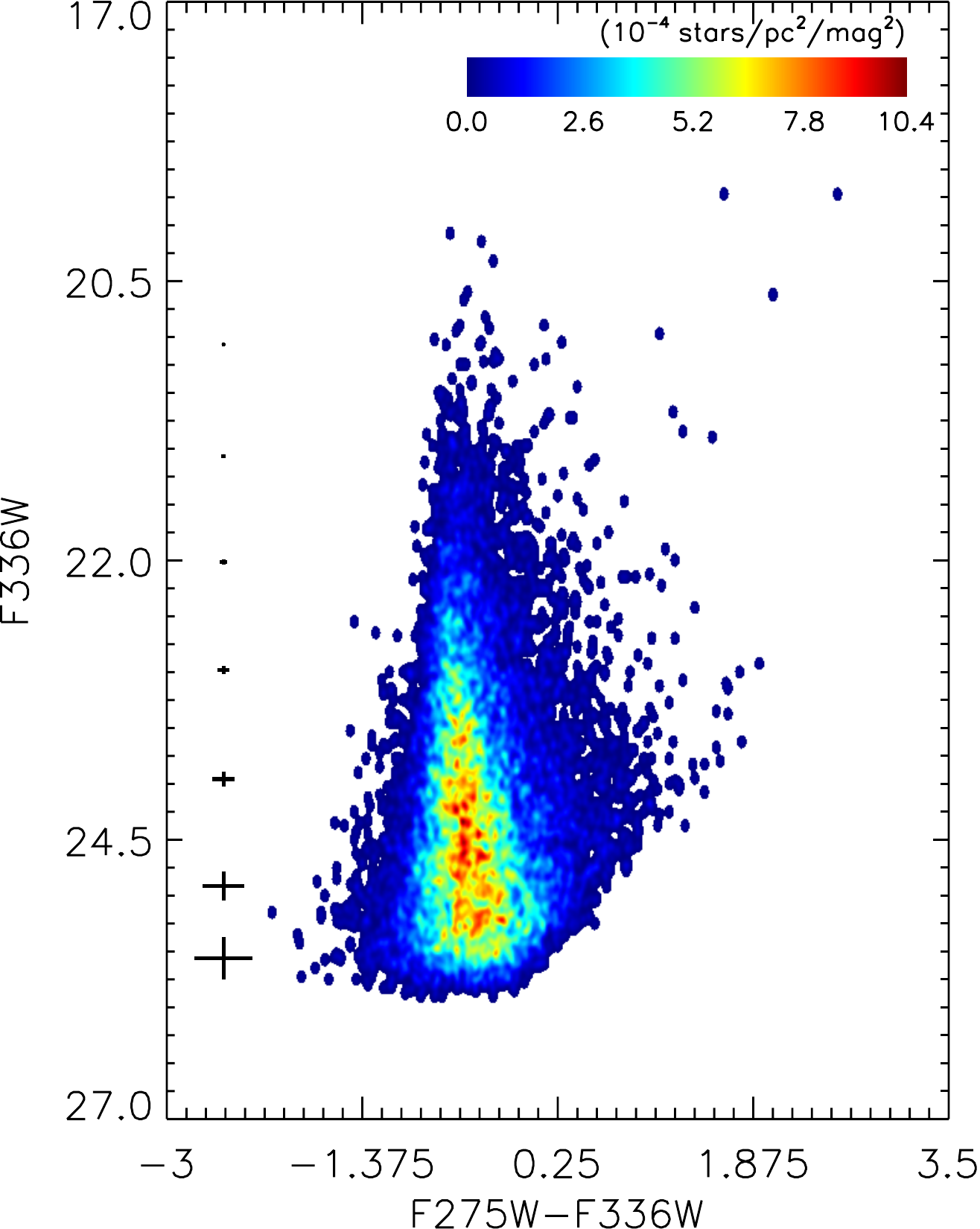} 
\caption{Hess diagram of stars identified in NGC\,6503 with the best photometric quality (see Sect.\,\ref{sect:selection}) 
in the LEGUS filter-pair F275W, F336W. Stellar surface densities across the Hess diagrams are 
indicated by the color bars. Crosses on the left represent typical photometric uncertainties. 
This sample, corresponding to the most recently formed stars
in NGC\,6503 is the `blue' sample in our analysis.
\label{fig:hess_275_336}}
\end{figure}

Photometry was performed with the package {\sc dolphot} \citep[e.g.,][]{dolphin00}.
This package performs point-spread function (PSF) fitting using PSFs especially
tailored to HST cameras. Before performing the photometry, we first prepared the 
images using the {\sc dolphot} packages {\sc acsmask} and {\sc splitgroups}. 
Respectively, these two packages apply the image defect mask and then split the 
multi-image STScI FITS files into a single FITS file per chip. We
then used the main {\sc dolphot} routine to make photometric measurements 
in each filter independently on the pre-processed images, relative
to the coordinate system of the drizzled reference image. The output 
photometry from {\sc dolphot} is on the calibrated VEGAMAG scale based 
on the zeropoints provided on the WFC3 page\footnote{\url{http://www.stsci.edu/hst/wfc3}}.
Indicative Hess diagrams of the complete stellar samples retrieved 
with this process are shown in Fig.\,\ref{fig:hess_diagrams}. From 
these diagrams it is seen that different wavelengths cover different stellar types,
with young stars in the blue filters and old stars in the red filters. 

{ It should be noted that 
our stellar photometry may include unresolved stellar clusters or multiple systems. 
However, the criteria applied in the following section for the selection of the best photometric 
measurements eliminate this contamination to the minimum. In any case, such systems, as well as 
unresolved binary systems, which should be still included in the best photometric sample, have a 
negligible effect to our analysis, because the measured blue light of every source is determined by 
that of the dominant bright star in the system.}

\begin{figure}
\centering
\includegraphics[width=\columnwidth]{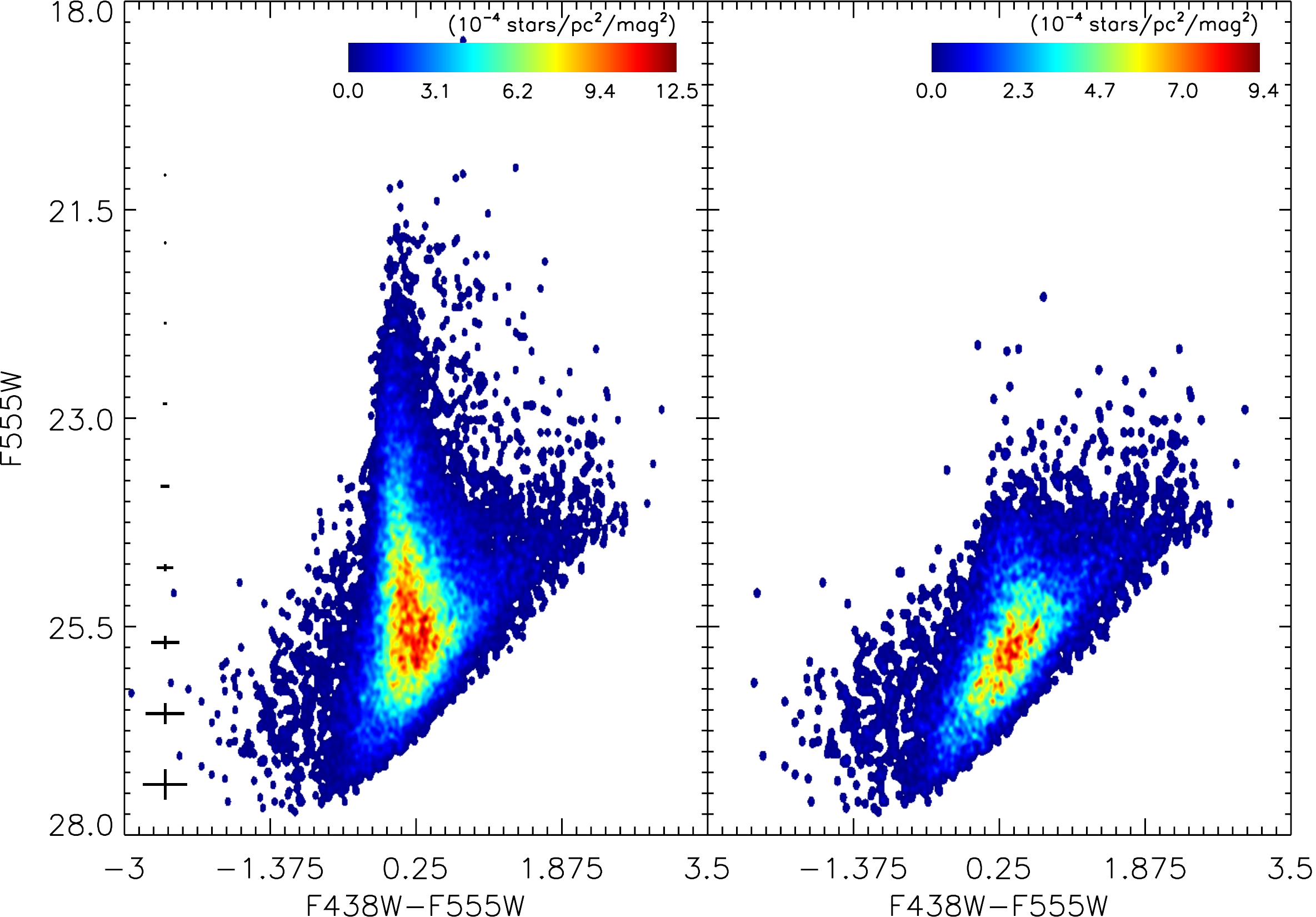} 
\caption{As in Figure\,\ref{fig:hess_275_336} for the stars identified with the best 
photometric quality in the LEGUS filter-pair F438W, F555W. 
{\em Left panel}: The Hess diagram of {\em all} stars detected in this filter-pair.  
{\em Right panel}: The Hess diagram of the remaining sources found {\em only} in F438W and F555W after 
 excluding those also observed in the F275W, F336W filter-pair. There is only a small fraction 
 of main-sequence stars included in this diagram, remained due to the longer exposures in these 
 filters than in F275W and F336W.
\label{fig:hess_438_555}}
\end{figure}

\begin{figure}
\centering
\includegraphics[width=\columnwidth]{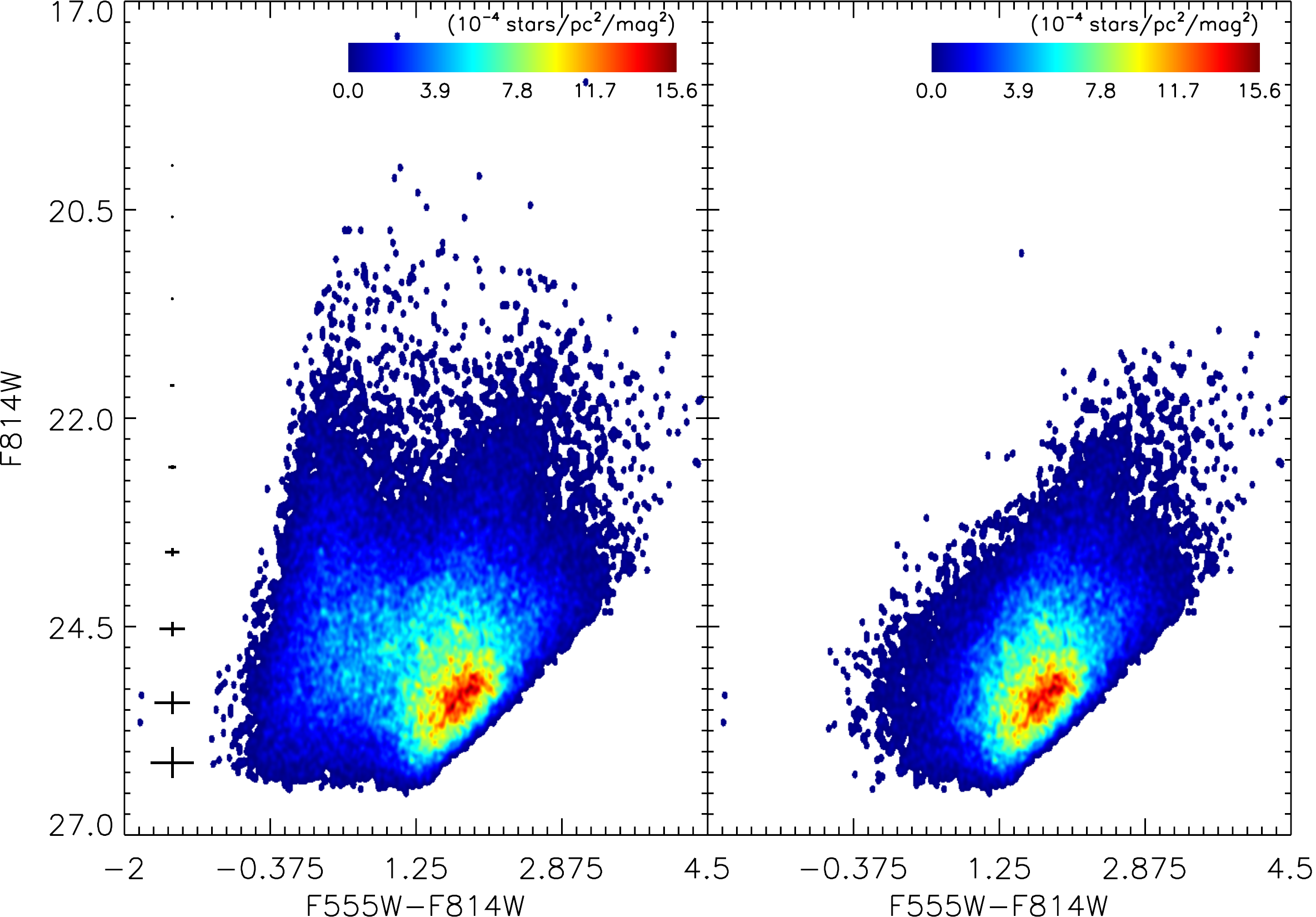} 
\caption{As in Figure\,\ref{fig:hess_438_555} for the stars identified 
with the best photometric quality in the LEGUS filter-pair F555W, F814W. Hess
diagrams are shown for stars detected {\em also} in the bluer filters (F275W, F336W, F438W; 
{\em Left panel}), as well as for those found {\em only} in the two considered bands ({\em Right panel}).
{ Fainter} main-sequence stars, which remained after the subtraction of 
sources identified also in the F275W and F336W filters, can be seen in the sample on the right. 
This residual population persisted due to the longer exposures in the optical filters than in
the UV filters. { The vast majority of the sources in the diagram on the right panel 
represents stars at later evolutionary stages. This is the `red' stellar sample in our analysis.} 
\label{fig:hess_555_814}}
\end{figure}

\subsection{Selection of the Stellar Sample}\label{sect:selection}


We separate the stellar sources with the most reliable photometry into four catalogs, 
each  including stars found in two adjacent filters throughout the complete wavelength 
coverage from F275W (UV) to F814W (I).  The waveband coverage of the filter-pairs overlap, 
so that, e.g., the bluer pair (F275W, F336W) shares filter F336W with the next redder 
pair (F336W, F438W), which shares filter F438W with the next and so forth. 
Since bluer filters cover the younger populations, while the redder ones identify mostly the 
more evolved stars, our selection distinguishes stars at roughly different evolutionary stages. 
These {\em best photometrically defined} stellar samples are determined in terms of quality parameters returned 
by our photometry. We apply the selection of this sample for stellar sources identified 
successfully in at least two adjacent filters. We compile the final photometric stellar 
catalogs in each filter-pair by applying the following quality criteria: 
\begin{itemize}
\item[] {\sc dolphot} type of the source, {\sc type}\,$=1$
\item[] Crowding of the source in each of the filters,  {\sc crowd}\,$< 2$ 
\item[] Sharpness of the source squared in each filter, {\sc sharp}$^2< 0.3$ 
\item[] Signal-to-noise ratio in each filter, {\sc snr}\,$> 5$ 
\end{itemize}

{ In {\sc dolphot}, the object type parameter has a value of 1 for the best stars in the photometry.
Stars too faint for PSF determination and non-stellar sources have  {\sc type}\,$>1$. 
The crowding parameter is a measure of how much brighter the star would have been measured had nearby stars not been fit simultaneously. For an isolated star it has the value of zero. The sharpness is zero for a perfectly-fit star, positive for a star that is too sharp, and negative for a star that is too broad\footnote{More details about the fit-quality parameters are given in {\sc dolphot} documentation, available at \href{http://americano.dolphinsim.com/dolphot/}{http://americano.dolphinsim.com/dolphot/}}.}

Statistics on each filter-pair stellar catalog are given in Table\,\ref{tab:photstat}. 
Stellar samples are given a color name (Col.\,1) based on the wavelength coverage of each filter-pair from blue 
to red (specified in Col.\, 2). Col.\,3 corresponds to the numbers of all stars found in the specific filter-pair, 
including sources detected also in the bluer filter-pair. We count how many more stars we detect in every redder filter-pair 
by selecting those stars that {\em have not been identified in the bluer filter-pairs}, and  in Col. 4 we provide the 
corresponding numbers of these ``additional'' stars. The corresponding percentages over the total numbers of stars in every 
filter-pair are given in Col. 5 of Table\,\ref{tab:photstat}.

\begin{table}
\centering
\caption{Statistics on stellar samples with best photometry in various filter pairs. Different 
stellar samples are coded by different colors, given in Col.\,1, throughout the text.}
\label{tab:photstat}
\begin{tabular}{llrrr}
\hline
\multicolumn{1}{c}{Sample} & \multicolumn{1}{c}{Filter} & \multicolumn{1}{c}{Total}  & \multicolumn{1}{c}{Number of} & \multicolumn{1}{c}{Percentage}  \\
                                                  & \multicolumn{1}{c}{pair}   & \multicolumn{1}{c}{number} & \multicolumn{1}{c}{sources not} & \multicolumn{1}{c}{over whole}  \\
                                                 &                                               & \multicolumn{1}{c}{of sources} & \multicolumn{1}{c}{in bluer filters} & \multicolumn{1}{c}{sample}\\
\hline \\
Blue & F275, F336 & 12834 &  & \\
Green & F336, F438 & 13499 & 1570&12\% \\
Yellow & F438, F555 & 21438 & 8484&40\% \\
Red & F555, F814 & 42511 & 27525&65\% \\
\hline
\end{tabular}
\end{table}

Indicative Hess diagrams of the blue (filter pair F275W, F336W), yellow (F438W, F555W) and red (F555W, F814W) stellar samples are shown in 
Figs.\,\ref{fig:hess_275_336}, \ref{fig:hess_438_555} and \ref{fig:hess_555_814}, respectively. In
the case of the two latter pairs, we show the Hess diagrams of both the complete stellar samples (left panels of Figs.\, 
\ref{fig:hess_438_555} and \ref{fig:hess_555_814}) and the stars remaining in the catalogs after subtracting those that 
have been identified also in the bluer filter-pairs (right panels of the figures). It is shown that after this subtraction, 
the majority of the main-sequence (MS) population is almost entirely eliminated in both diagrams, indicating more 
clearly that they correspond mostly to evolved stellar populations. Small contamination of the blue-faint part of these diagrams 
by residual MS stars is expected due to the deeper observations in the corresponding filters in comparison to the 
blue filters and due to photometric uncertainties.

\begin{figure*}
\centering
\includegraphics[width=0.85\textwidth]{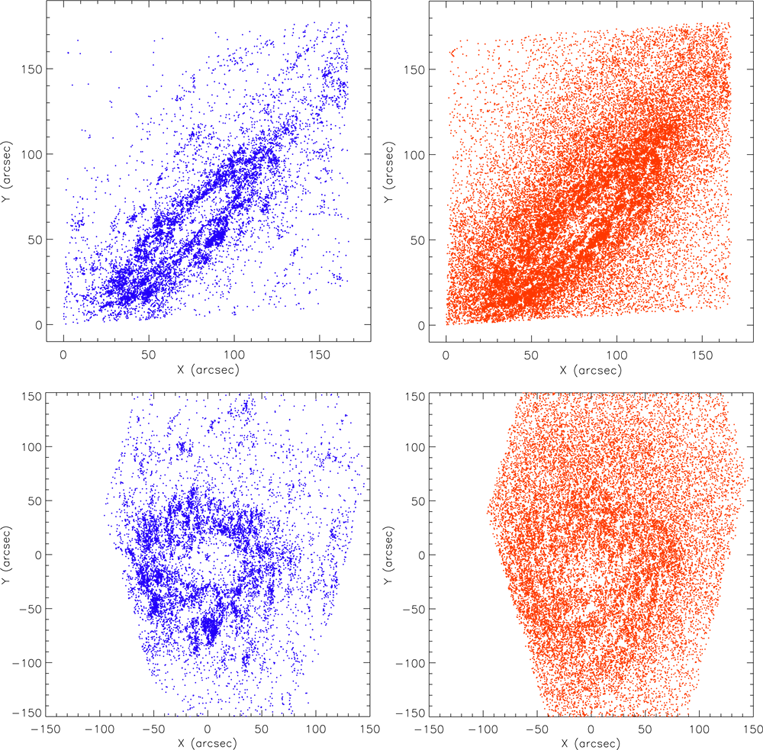} 
\caption{{\em Top panels} -- Spatial distribution of stars detected in two filter-pairs with the best photometry. 
{\em Left panel}: Stars in the sample detected in the F275W and F336W filters (``blue'' sample). 
{\em Right panel}: Stars detected in the filter-pair F555W and F814W (``red'' sample). 
{\em Bottom panels} -- The same distributions deprojected for the inclination of the galaxy (see Section\,\ref{s:deproject}).
The distribution of ``red'' sources is more extended than the ``blue'' sources, consistent with the prolonged dynamical mixing 
of these sources. The overall { distribution} of the red sources, however, still tracks partly star-forming regions of the galaxy.
Gaps in the populated central areas of the red stellar distribution, { and the lack of evidence of a central bar}, are partially assigned to photometric incompleteness,
while the lack of blue stars represents a true deficiency at the galaxy center (see Sect.\,\ref{s:ssdp}).
\label{fig:maps-arcsec}}
\end{figure*}




The comparative study of different stellar samples, as selected above, allows us to investigate 
the differences in the clustering behavior between stars that correspond to different evolutionary 
stages, from the most recent star formation events to those corresponding to earlier stellar 
generations in NGC\,6503. Our investigation deals with the clustering behavior of the most recent
stellar generation in the galaxy, i.e., the blue sample (Fig.\,\ref{fig:hess_275_336}), but for 
comparison we also address the distribution of the general population of NGC\,6503, 
by considering stars detected in the red sample (Fig.\,\ref{fig:hess_555_814} -- right panel). 
Indeed, stars in these two evolutionary stages demonstrate different distributions across the observed field-of-view. 
This is shown, for example, in the map of the stars in the blue sample (found in the F275W and F336W filters) and that 
of the red stellar sample (found in the F555W and F814W filters), displayed in Fig.\,\ref{fig:maps-arcsec} (top panel). 
The young blue stars  appear to populate more well-defined structures, while the red population is much more extended, populating 
the whole observed disk of the galaxy. 

We compared these distributions against the light distribution from 
{\em Spitzer} images from \cite{dale09} in 4.5\,$\mu$m, tracing old stars, and 8\,$\mu$m, 
an indicator based on dust emission of where the young stars are, in order to check whether the 
observed stellar distributions could be affected by dust attenuation. This comparison confirmed that  the blue stars 
are more clustered in a ring-like structure, traced by the 8\,$\mu$m image, 
{ while the 4.5\,$\mu$m image shows a smooth distribution for the old populations}\footnote{ It is interesting to note that the 4.5\,$\mu$m image shows only a faint signature of a bar.}.
Nevertheless, the distributions shown in Fig.\,\ref{fig:maps-arcsec} (top panel) suffer from projection effects that 
limit our analysis on the global stellar clustering through the observed stellar distributions. Thus, we correct
 the observed stellar positions for the known inclination of the galaxy. 

\begin{figure*}
\centering
\includegraphics[width=1.\textwidth]{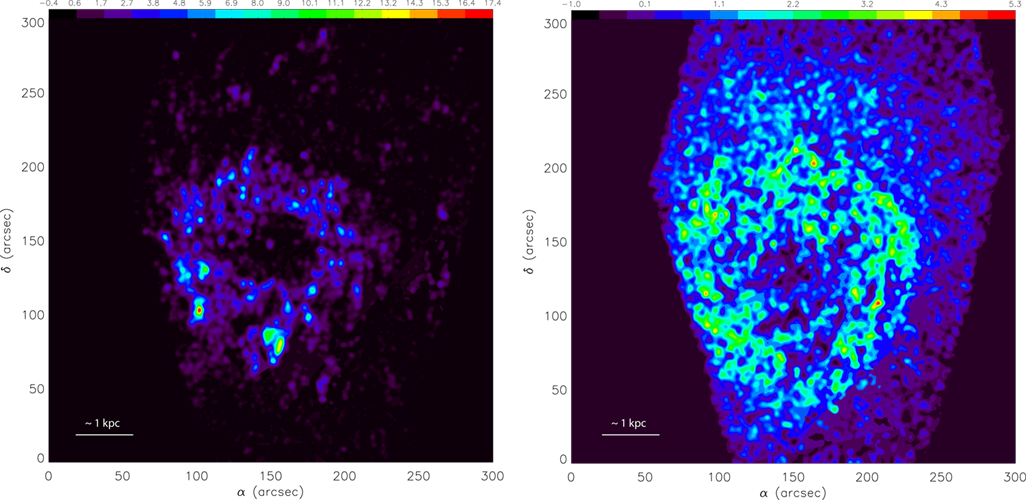} 
\caption{Surface stellar density maps constructed with the {\em Kernel Density Estimation} method with a kernel of FWHM\,$\sim$\,80\,pc, 
for stars identified with the best photometry in the filter pairs F275W, F336W (left panel), and F555W, F814W (right panel). The color bar at 
the top corresponds to stellar density significance levels in values of $\sigma$, where $\sigma$ is the density standard deviation in the whole 
of each sample.  The `blue'  stellar 
population surface density tracks very well the star-forming ring of NGC\,6503 \citep[e.g.,][]{mazzuca08, freeland10}. LEGUS
photometry reveals individual star-forming structures across the ring in order to access the large-scale progression of star formation in space and time. 
On the other hand, the `red' evolved stellar population is far more extensively distributed than the blue, populating the whole disk of the
galaxy. This difference is consistent with the more prolonged dynamical mixing of more evolved populations, and their dissolution in the field of NGC\,6503. \label{fig:kde_maps}}
\end{figure*}

\subsection{Correction for projection of the galactic disk\label{s:deproject}}

In order to assess the spatial distribution of stars, free from the effect of projection of the galactic disk,
we apply a deprojection of the stellar positions from the plane of the sky to the plane of the galaxy, under 
the simple assumption of an axisymmetric flat rotating disk for NGC\,6503. This assumption is based on
the remarkably regular gas kinematics in NGC\,6503 that are well described by rotation only \citep{kuziodenaray12}.

For a star with position (x, y) in the plane of the sky (where the center of the galaxy is at the origin), a simple rotation 
by the position angle, $\phi$, yields
\begin{eqnarray}
          \begin{array}{r@{\,\,}l}
          x^\prime= & x\,{\rm sin}(\phi) - y\,{\rm cos}(\phi)\\
          y^\prime= & x\,{\rm cos}(\phi) + y\,{\rm sin}(\phi).
                    \end{array}
\label{eq:deproject}
\end{eqnarray}
The coordinates of the star in the mid-plane of the galaxy are then 
$x^\prime$ and $y^\prime/ {\rm cos}( i)$, $i$ being the inclination angle.
We rotated our stellar catalogs for a position angle of $\phi = 135^\circ$,
determined by the orientation of the observed field-of-view. We then corrected
them for projection for an inclination angle of $i=$\,75.1$^\circ$, over its kinematic 
center (J2000) 17$^{\rm h}$49$^{\rm m}$26.30$^{\rm s}$, 
70$^\circ$08$^\prime$40.7\arcsec \citep[both determined by][]{greisen09}.
The  spatial distributions of the blue and red stellar samples 
corrected for projection are shown in Fig.\,\ref{fig:maps-arcsec} (bottom panel).
It is interesting to note that the deprojected distribution of the blue young stellar 
population (Fig.\,\ref{fig:maps-arcsec} -- bottom-left panel) is in remarkable 
agreement of the deprojected GALEX NUV data, presented by \cite{freeland10}.
Both distributions highlight the star-forming ring of the galaxy and its nucleus LINER  
\citep[low-ionization nuclear emission-line regions;][]{lira07}, 
shown as a UV-brigh clump at the center of the ring \citep[][their Fig.\,5]{freeland10}.  
In the following section we further compare the distribution of our blue stellar sample with that of the red sample, through 
their stellar surface density maps.

\subsection{Stellar Surface Density Maps}\label{s:method}

We construct stellar surface density maps for the blue and red samples with the application 
of the {\em Kernel Density Estimation} (KDE), by convolving the stellar catalogs with a Gaussian 
kernel. The FWHM of this kernel specifies the ``resolution'' at which stellar structures will be revealed. 
These density maps can be treated as significance maps, corresponding to two dimensional 
probability functions for the positions of the stars in each sample. For their construction with this 
method, the most important (and debatable) parameter is the size of the KDE kernel 
(the convolving FWHM), i.e., the resolution of the constructed  maps. At the 
distance of NGC\,6503 \citep[$\sim$\,5.3\,Mpc;][]{karachentsev03}, 1 second of arc corresponds 
to $\sim$\,25.7\,parsec, and therefore every WFC3 UVIS pixel ($\simeq$\,0.04\arcsec) corresponds to a physical scale 
of $\sim$\,1\,pc. Consequently a KDE map constructed from our data with a resolution of $\sim$\,10\,pc
requires a kernel with FWHM of $\sim$\,10\,pixels ($\sim$\,0.4\,arcsec). However, such a map would be
extremely noisy, revealing small compact stellar over-densities rather than coherent physically related stellar concentrations. 
The KDE kernel size to be applied depends on the science to be achieved from the KDE maps, and it is best decided based on 
experimentation. Our tests on various kernel sizes showed that a FWHM of $\sim$\,80\,pc, comparable to the scale of typical 
OB associations \citep[][and references therein]{gouliermis11} and giant molecular clouds \citep[GMCs;][and references therein]{bolatto08}, 
is the most appropriate for the detection of large star-forming structures. 

The stellar surface density maps of NGC\,6503, constructed from our photometry with a kernel of $\sim$\,80\,pc 
(80 UVIS pixels), for both the blue and red stellar samples are shown in Fig.\,\ref{fig:kde_maps}. 
From the KDE map of the blue sample (Fig.\,\ref{fig:kde_maps} -- left panel) it is seen that the young blue stellar population 
depicts the inner star-forming ring of NGC\,6503  \citep[e.g.,][]{mazzuca08, freeland10}, where individual large structures 
can be identified. On the other hand, the red stellar sample found in filters F555W and F814W is much more 
spread out, populating the whole extent of the disk of the galaxy.
In the following sections we further explore the clustering behavior of young stars in NGC\,6503. Specifically, we 
(a) conduct a census of young stellar structures, and determine their demographics, across the extent of the observed field-of-view, and 
(b) investigate the global distribution of the young populations and assess its hierarchy in order to understand the global star formation 
topology in NGC\,6503.


\section{Young Stellar Structures in NGC\,6503}\label{s:structident}

The blue stellar population selected in the previous sections for the study of 
clustered star formation comprises 12\,834 stars. Distinct concentrations of these 
young stars can be identified as stellar over-densities in the KDE stellar surface density maps. 
These over-densities correspond to individual star-forming structures, such as stellar associations, stellar aggregates and 
stellar complexes, the large-scale centres of star formation in a galaxy. The identification and characterization 
of these systems is thus important to our understanding of how galaxies construct their stellar components and
what is their future evolution. 
In this section we compile a detailed catalog of all young stellar concentrations that
can be revealed from our observations of NGC\,6503. Our method is applied by 
first constructing the stellar surface density KDE map of Fig.\,\ref{fig:kde_maps}, 
and by identifying young stellar structures  as stellar over-densities at specific significance levels, 
measured in $\sigma$ above the average density, $\sigma$ being the 
standard deviation of the background density \citep[see, e.g.,][for more details on the method]{gouliermis00, gouliermis10}.

Independent repetitions of the detection method for different density thresholds, i.e. different 
significance levels, reveals young stellar concentrations across the complete dynamic range in
stellar density.  This process reveals smaller, denser stellar concentrations systematically 
belonging to larger and looser ones, providing the first evidence of hierarchical structural morphology 
in the distribution of the blue stars in NGC\,6503. We refer to all these concentrations generally as 
{\em stellar structures}. 
In order to eliminate projection effects in our detection and the determination of the morphological 
parameters of the detected structures we apply our method on projection corrected maps, such as those
shown in Fig.\,\ref{fig:kde_maps}.


\begin{table*}
\centering
\caption{Survey of the young stellar structures in NGC\,6503. Explanations on the characteristic parameters 
of the structures are given in the text (Sect.\,\ref{s:methclus}). Only the first 15 records of the survey are shown here for reference. The complete catalog 
of 244 structures is available on-line at LEGUS site \href{https://legus.stsci.edu/}{https://legus.stsci.edu/}.}
\label{tab:cluslist}
\begin{tabular}{rrrcrrrcccccccc}
\hline
{ID} & 
{$\sigma$} &
{R.A.} &
{Decl.} &
\multicolumn{1}{c}{$N_\star$} &
\multicolumn{2}{c}{Radius (pc)} &
{$\xi$} &
{$\rho$ ($\times 10^{-3}$)} &
{275$_{\rm tot}$}  &
{336$_{\rm tot}$}  &
{438$_{\rm tot}$} &
Family &
Group \\
{} & 
{} &
\multicolumn{2}{c}{J2000 (deg)} &
{} &
\multicolumn{1}{c}{$r_{\rm eff}$} &
\multicolumn{1}{c}{$r_{\rm max}$} &
{} &
{(stars\,\,pc$^{-2}$)} &
(mag) &
(mag) &
(mag) &
id.&
id.\\
\hline 
         1&           1&  267.366870&   70.139878&        4388&      1158.7&      2424.7&       2.093&        1.04&       13.73&       14.21&       15.13&     1&  ~~1\\
         2&           1&  267.360523&   70.146816&        2189&       846.0&      1572.5&       1.859&        0.97&       14.68&       15.13&       16.02&     2&     21\\
         3&           1&  267.399790&   70.138429&         643&       452.4&       670.2&       1.481&        1.00&       16.26&       16.67&       17.57&     3&     36\\
         4&           1&  267.324116&   70.151839&         206&       255.9&       353.4&       1.381&        1.01&       17.12&       17.30&       17.93&     4&     37\\
         5&           1&  267.315085&   70.150455&         114&       200.6&       375.2&       1.871&        0.91&       18.18&       18.55&       19.43&     5&     38\\
         6&           1&  267.360586&   70.148180&          97&       194.2&       252.0&       1.298&        0.83&       18.26&       18.74&       19.43&     6&      39\\
         7&           1&  267.317336&   70.151773&          77&       176.4&       238.8&       1.354&        0.80&       18.92&       19.26&       20.13&     7&       40\\
         8&           1&  267.325148&   70.143415&          62&       148.4&       212.8&       1.434&        0.91&       18.58&       19.02&       19.78&     8&       41\\
         9&           2&  267.392443&   70.135681&        1197&       481.5&       840.8&       1.746&        1.64&       14.94&       15.44&       16.46&     1&      ~~1\\
        10&           2&  267.349994&   70.141340&        1007&       419.9&      1022.3&       2.435&        1.82&       15.09&       15.62&       16.57&   1&      ~~2\\
        11&           2&  267.399997&   70.138375&         529&       362.4&       547.9&       1.512&        1.28&       16.49&       16.90&       17.78&     3&     36\\
        12&           2&  267.345413&   70.149136&         515&       357.1&       751.0&       2.103&        1.29&       16.23&       16.71&       17.54&     2&     21\\
        13&           2&  267.365176&   70.140677&         319&       278.0&       551.5&       1.984&        1.32&       16.57&       17.05&       17.87&     1&      ~~3\\
        14&           2&  267.376002&   70.145055&         319&       261.3&       604.3&       2.313&        1.49&       16.78&       17.28&       18.19&     2&     22\\
        15&           2&  267.333935&   70.151913&         181&       205.8&       391.5&       1.902&        1.37&       17.49&       17.84&       18.78&     2&     23\\
\hline
\end{tabular}
\end{table*}

\begin{figure*}
\centering
\includegraphics[width=\textwidth]{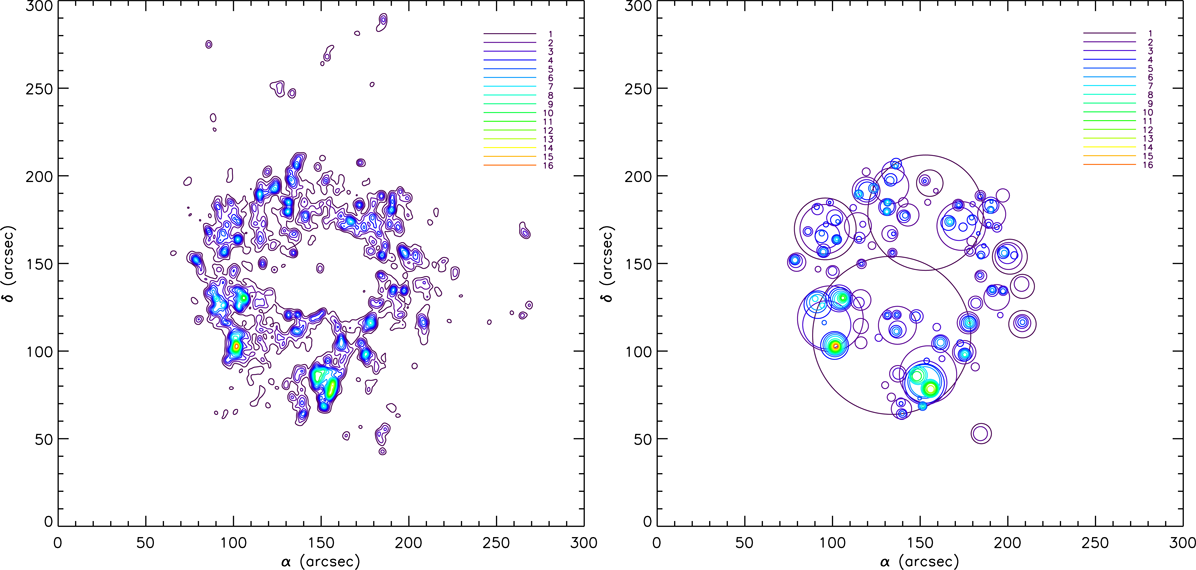} 
\caption{{\em Left panel.} Iso-density contour plot of the surface stellar density map constructed with the KDE method with a kernel of 
FWHM\,$\sim$\,80\,pc, for the blue stars identified in both filters F275W and F336W. {\em Right panel.} Chart of the identified 
stellar structures, i.e., those which meet the criteria of including $\geq$\,5 stellar members, and appearing in at least two
surface density significance levels. Systems borders are drawn with circles { representing their sizes, $r_{\rm eff}$}, as determined at the various 
significance levels, where they appear. In both panels different colors are used for the isopleths and apertures of systems found at 
different significance levels in steps of 1$\sigma$.
\label{fig:blue_kde_map}}
\end{figure*}

\subsection{Detected Young Stellar Structures}\label{s:methclus}

Young stellar structures are identified on the KDE stellar surface density map of the blue stars. 
With 80\,pc resolution we are able to identify loose stellar concentrations that correspond to the stellar associations and stellar complexes of the galaxy, 
while avoiding a high noise level produced by random small-scale density fluctuations. 
An iso-density contour plot of this surface density map is shown in Fig.\,\ref{fig:blue_kde_map} (left panel). In this plot, contour lines 
correspond to different density levels, starting from that corresponding to the average density (0$\sigma$) up to the 
highest level of 16$\sigma$. They are drawn in steps of 1$\sigma$ with different colors accordingly. 
These {\em isopleths}\footnote{The noun ``isopleth'' is used to define every line on the map that connects points having 
equal numeric value of surface stellar density; Origin from the Greek {\em ``isopl\={e}th\={e}s''}, equal in number.} 
determine the borders of various stellar structures identified across NGC\,6503 at various significance levels. 
The physical dimensions of each detected structure are defined by these borders. A parameter considered 
in our detection is the minimum number of stars included in an over-density in order to be classified as a structure, 
which is $N_{\rm min} = 5$, in line with other identification techniques \citep[e.g.,][]{bastian07}.
This criterion eliminates the detection of random stellar congregations, so-called asterisms.  
Additionally, we consider as real those structures which appear in at least two significance levels, while we 
treat features that appear in only one (essentially the lowest 1$\sigma$) level as spurious detections. This criterion provides 
confidence that the identified over-density is indeed a physical stellar concentration.


\begin{figure}
\centerline{\includegraphics[clip=true,width=0.475\textwidth]{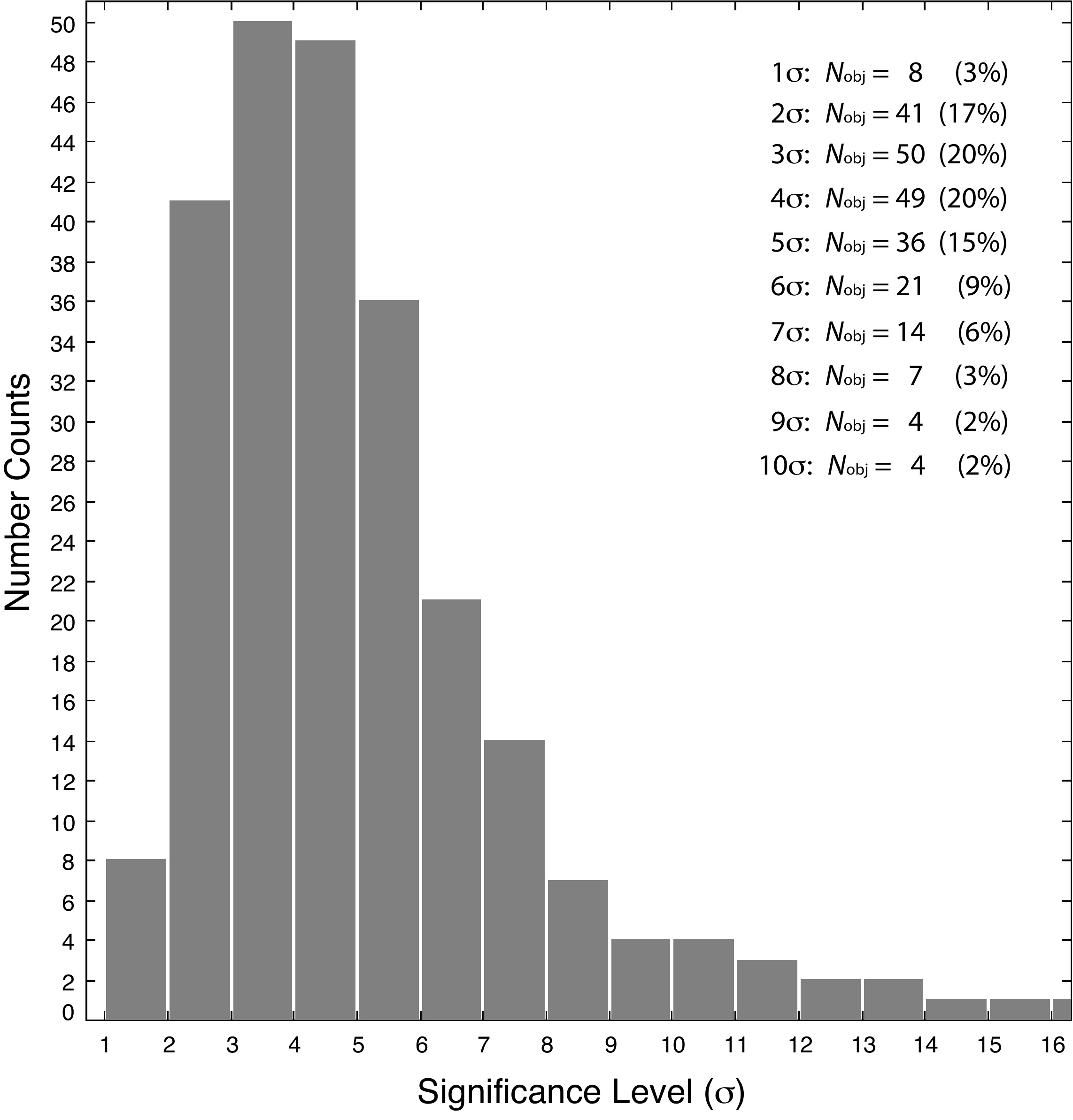}}
\caption{Histogram of the numbers of detected young stellar structures per significance level in NGC\,6503.
The majority of the structures are revealed at the 2, 3 and 4$\sigma$ levels. They represent the 
population of stellar aggregates and complexes  of the galaxy. All detected structures are members of
eight extented  structures found at the 1$\sigma$ level. Three of them are large super-structures encompassing 
95\% of the total structures population.
 \label{f:str_stat_lev}}
\end{figure}

The catalog of identified stellar structures in NGC\,6503 consists of 244 systems, revealed at significance levels between 1 and 16$\sigma$. In Table\,
\ref{tab:cluslist} we show a sample of the first 15 records of this catalog and their measured 
parameters. The complete catalog of all identified young stellar structures is available on-line. In columns 1 and 2 the ID number and the detection density 
threshold (in $\sigma$) are given for every object. Columns 3 and 4 provide the celestial coordinates of 
the structures' barycentres, which correspond to their KDE density centres. Column 5 shows the number of the blue stars included 
within the borders of every structure, as defined by the corresponding isopleth. A measure of the size of each system is the so-called {\em effective radius} 
\citep[e.g.,][]{carpenter00} or {\em equivalent radius} \citep[e.g.,][]{romanzuniga08}, defined as the radius of a circle with the same area as the 
area covered by the system. We provide two measurements for this radius: (1) The radius determined by the area $A_{\rm CH}$ enclosed by the 
convex hull of the system ($r_{\rm eff} \equiv \sqrt{A_{\rm CH}/\pi}$, Col.\,6) and (2) the radius defined by the area $A_{\rm max}$ enclosed by the largest circle 
that encompass the entire system ($r_{\rm max} \equiv \sqrt{A_{\rm max}/\pi}$, Col.\,7), equivalent to the half of the distance between the two furthest 
sources in the system. 

The ratio of these two radii provides a characterization of the {\em elongation} of each system $\xi \equiv r_{\rm max} / r_{\rm eff}$, which we provide in 
column 8. \cite{schmejaklessen06} showed that the estimation of the system's area from its convex hull makes $\xi$ a reliable measure for the elongation, 
since it excludes the possibility that fractal substructure in an otherwise spherical structure can lead to unrealistic elongations. These authors also determined 
the increase of $\xi$ with increasing axis ratios of elliptical distributions; A circular distribution (axis ratio\,$=$\,1) has $\xi \simeq 1$. From the number of blue 
stars enclosed within the borders of each structure (Col. 5) and its radius $r_{\rm eff}$ (Col. 6) we also have a measurement of the surface stellar 
density of each structure, which is given in column 9 of Table\,\ref{tab:cluslist}. These stellar structures contain blue stars in the whole observed magnitude 
range of $F336W$~\lsim~26. The total brightness of each system, calculated from its stellar population in the three blue filters (i.e., F275W, F336W and F438W) is 
given in columns 10, 11 and 12 respectively.

A histogram of the numbers of all detected structures per detection density level up to 10$\sigma$ 
is shown in Fig.\,\ref{f:str_stat_lev}.  This distribution peaks at the detection levels of 2, 3 and 4$\sigma$,
which determine the vast majority of the detected structures. Higher detection levels mostly reveal 
the high-density peaks of the same structures, as well as their ``offspring'' structures which they break 
into. A chart of the detected structures is shown in Fig.\,\ref{fig:blue_kde_map} (right panel). In this map the structures are indicated by 
circles equivalent to their measured sizes. Apertures with radii equal to $r_{\rm eff}$ of the detected systems are drawn in 
different colors according to the density threshold where the structures were detected (as in the isopleth plot at the left 
panel of the figure). 

As shown in Table\,\ref{tab:cluslist}, and can be seen in the maps of Fig.\,\ref{fig:blue_kde_map}, there are eight 
large structures identified at the 1$\sigma$ level. Two of them extend to the 2$\sigma$ level, and appear isolated, i.e., not related to 
other large structures.  Three additional 1$\sigma$ structures appear also in few intermediate density levels ($\geq$\,2$\sigma$), two 
of them also showing a second sub-structure.  The remaining three 1$\sigma$ structures are extended  stellar concentrations 
that qualify as large {\em stellar super-complexes} with dimensions between 1 and 2\,kpc. These structures comprise the vast majority  
of identified stellar structures, which are themselves {\em multiple} concentrations seen at higher density levels. The 2$\sigma$ 
concentrations most probably correspond to the so-called {\em stellar aggregates} and {\em stellar complexes} of the galaxy, while 
the structures detected at the 3$\sigma$ level are actually members of these concentrations, fulfilling the typical image of hierarchical 
structuring of stars on a galaxy-scale. Even higher density detections ($\geq$\,4$\sigma$) correspond to the more condensed stellar 
density peaks of the larger structures.

We define each of the structures revealed at the 1$\sigma$ level as a {\em family} of structures. We also determine as members of  
a {\em group} structures that are ``spawned'' from the same single structure. In total we identify 82 groups divided into eight families, 
three of which are large super-complexes that encompass 95\% of all revealed stellar structures. The family and group numbers of 
each structure are given in Cols. 13 and 14 of Table\,\ref{tab:cluslist} respectively.

\begin{figure}
\centerline{\includegraphics[clip=true,width=0.475\textwidth]{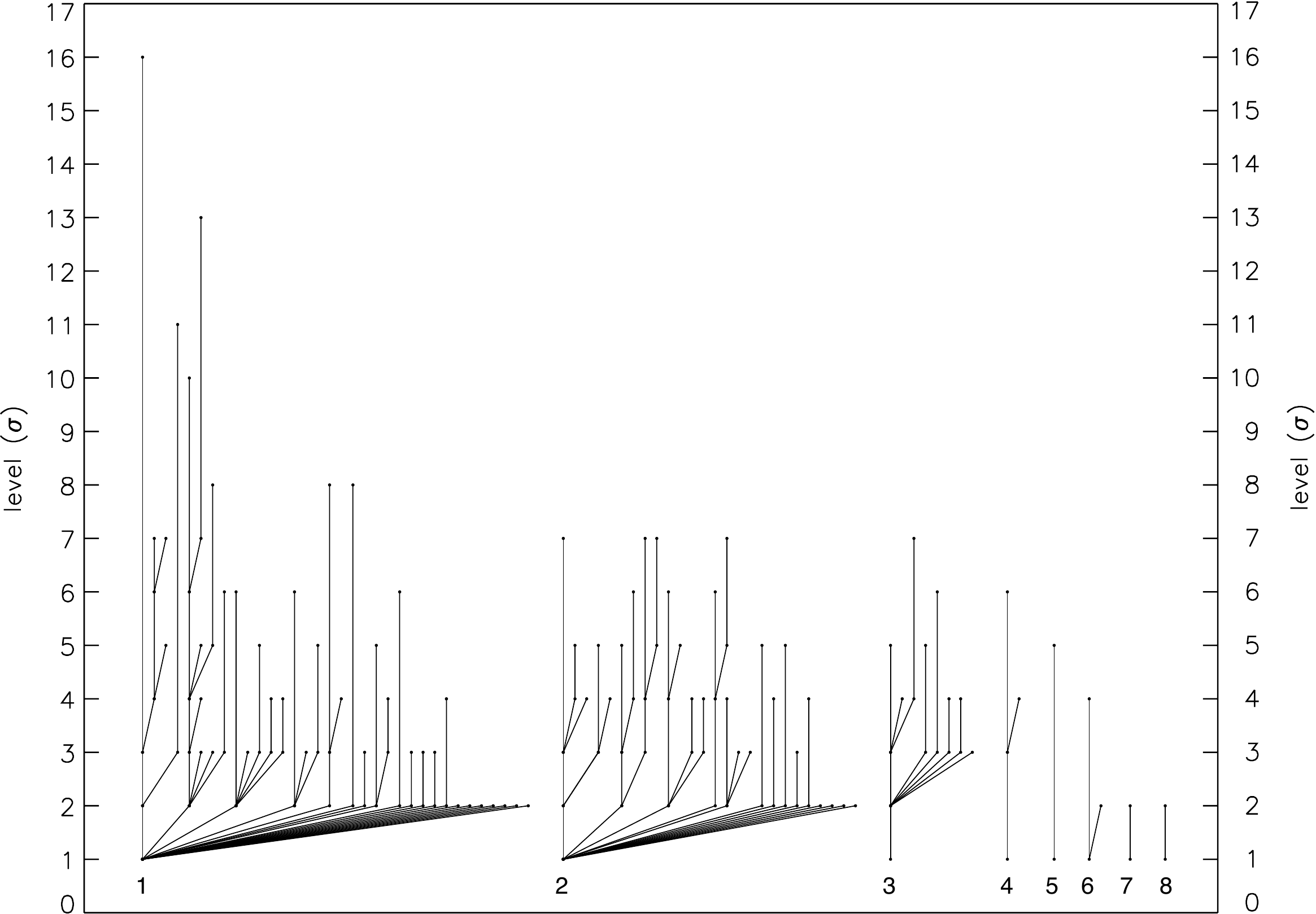}}
\caption{Dendrogram (structure tree) of the stellar structures identified at different significance levels in the 
KDE density map (Fig.\,\ref{fig:blue_kde_map}), illustrating the hierarchy in the manner stellar structures of 
blue stars are assembled. This dendrogram also demonstrates the hierarchical behavior in the global clustering of
these stars across the whole galaxy, which we quantify in more detail in Section\,\ref{s:ACF}. Parental super-structures,
identified at 1$\sigma$ level, are indicated by their identification numbers.
\label{f:dendrogram}}
\end{figure}

\begin{table*}
\centering
\caption{Demographics of the young stellar structures revealed in the KDE surface stellar density map at various significance levels. In Col. 1 the detection levels 
(in $\sigma$) are given.  
The parameters shown are the average size (Col.\,2) and size dispersion (Col.\,3), the average elongation  (Col.\,4), and the average surface stellar density  (Col.\,5) of
the structures in each density level. The corresponding total stellar numbers ($N_{\star}$, Col.\,6) and UV magnitudes ($m_{\rm 275}$, Col.\,8) per density level are also given, along with the corresponding fractions of these parameters over the total observed number of stars, $N_{\star, {\rm tot}}$, and total observed stellar UV flux, $m_{\rm 275, tot}$, shown in Cols.\,7 and 9, respectively. There are no dispersions given for the parameters in the last three entrances, because there is 
only one object found in each of the corresponding density levels.}
\label{tab:structdemo}
\begin{tabular}{crrccrccc}
\hline
\multicolumn{1}{c}{Detection} &
\multicolumn{2}{c}{Size $S$ (pc)} &
\multicolumn{1}{c}{$\bar{\xi}$} &
\multicolumn{1}{c}{$\bar{\rho}$ ($\times 10^{-3}$)} &
\multicolumn{1}{c}{$N_{\star}$} &
\multicolumn{1}{c}{$f_{\star}$}&
\multicolumn{1}{c}{$m_{\rm 275}$}&
\multicolumn{1}{c}{$f_{\rm UV}$}\\
\multicolumn{1}{c}{level ($\sigma$)} &
\multicolumn{1}{c}{$\bar{S}$} &
\multicolumn{1}{c}{$\sigma_{S}$} &
\multicolumn{1}{c}{}&
\multicolumn{1}{c}{(stars\,\,pc$^{-2}$)}&
\multicolumn{1}{c}{}&
\multicolumn{1}{c}{$N_{\star}/N_{\star, {\rm tot}}$}&
\multicolumn{1}{c}{(mag)}&
\multicolumn{1}{c}{$m_{\rm 275}/m_{\rm 275, tot}$}\\
\multicolumn{1}{c}{(1)} &
\multicolumn{1}{c}{(2)} &
\multicolumn{1}{c}{(3)} &
\multicolumn{1}{c}{(4)}&
\multicolumn{1}{c}{(5)}&
\multicolumn{1}{c}{(6)}&
\multicolumn{1}{c}{(7)}&
\multicolumn{1}{c}{(8)}&
\multicolumn{1}{c}{(9)}\\
\hline 
   1&    858&    751&   $1.60\pm 0.30$&$   0.93\pm 0.08$&   7776&    0.606&   13.2&   0.699\\
   2&    295&    210&   $1.49\pm 0.39$&$   1.46\pm 0.20$&   6126&    0.477&   13.3&   0.603\\
   3&    202&    132&   $1.32\pm 0.29$&$   1.97\pm 0.26$&   4371&    0.341&   13.6&   0.474\\
   4&    148&    104&   $1.30\pm 0.33$&$   2.94\pm 1.00$&   3070&    0.239&   13.9&   0.367\\
   5&    127&     94&   $1.17\pm 0.20$&$   3.46\pm 0.84$&   2043&    0.159&   14.2&   0.283\\
   6&    129&     97&   $1.22\pm 0.28$&$   3.86\pm 1.04$&   1400&    0.109&   14.5&   0.211\\
   7&    121&     88&   $1.18\pm 0.26$&$   4.80\pm 1.54$&    907&    0.071&   14.8&   0.151\\
   8&    142&     92&   $1.21\pm 0.29$&$   5.20\pm 2.11$&    623&    0.049&   15.1&   0.117\\
   9&    179&     45&   $1.34\pm 0.19$&$   4.55\pm 0.52$&    475&    0.037&   15.3&   0.099\\
  10&    146&     47&   $1.31\pm 0.23$&$   5.15\pm 0.90$&    362&    0.028&   15.5&   0.082\\
  11&    125&     63&   $1.32\pm 0.39$&$   6.08\pm 0.80$&    243&    0.019&   15.8&   0.063\\
  12&    138&      6&   $1.47\pm 0.48$&$   6.05\pm 0.60$&    179&    0.014&   16.0&   0.052\\
  13&    110&     10&  $ 1.57\pm 0.56$&$   6.43\pm 0.56$&    121&    0.009&   16.2&   0.044\\
  14&     99&   -   &   $1.14$&$   7.50  $&     57&    0.004&   16.8&   0.024\\
  15&     81&   -   &   $1.24 $&$   8.99   $&     45&    0.004&   17.0&   0.021\\
  16&     59&   -   &   $1.07  $&$   6.87   $&     18&    0.001&   17.4&   0.014\\
\hline
\end{tabular}
\end{table*}

\subsection{Structure Tree}\label{s:structtree}

In this section we 
visualize the ``family ties'' of high density structures in relation to their parental super-structures. 
The ``breaking'' of super-structures of specific low density level into several, smaller and denser sub-structures at higher density levels, 
provides the first evidence of a morphological hierarchy in the way young stars are clustered across the ring.  An intuitive way to illustrate hierarchical structures 
is through the so-called {\em dendrograms}, introduced as ``structure trees''  for the analysis of molecular cloud structure by \cite{houlahan92}, 
refined by \cite{rosolowsky08}. A dendrogram is constructed by identifying in the stellar surface density map connected structures found at different 
density thresholds, while keeping track of the connection to ``parent structures'' (on a lower level) and ``child structures'' 
(on the next higher level, lying within the boundaries of its parent). In the dendrogram of a geometrically perfect hierarchy  
each parent would branch out into the same number of children at each level.

We construct the dendrogram of the stellar structures detected in NGC\,6503 at various density thresholds, up to the 
highest level of 16$\sigma$ above the background density. In this dendrogram, 
shown in Fig.~\ref{f:dendrogram}, the structures found at each density level  are represented not only by the `leaves' 
that end at the particular level, but by all branches present  at that level. For example, at the 3$\sigma$ level, there 
are 50 branches of the  dendrogram (regardless whether they end at this level, continue to a higher level as a single 
system, or split into two or more branches), corresponding to the 50  detected stellar structures. This dendrogram 
demonstrates that most structures split up into several sub-structures over few levels.
The combination of this dendrogram with the maps of Fig.~\ref{fig:blue_kde_map} illustrates graphically 
the hierarchical spatial distribution of young stars in NGC\,6503.

\subsection{Parameters demographics}\label{s:params}

Our survey of young stellar structures (Table~\ref{tab:cluslist}) covers a variety of systems, starting with those including the minimum of 6 to 8 stellar members with dimensions 
\lsim\,40\,pc (e.g., structures\,Ê148 and 226), up to those with maximum size of more than 1\,kpc, including 
over 1000 stars (i.e., structure 9). 
The parameters demographics of the stellar structures, revealed at various significance levels, are given in Table\,\ref{tab:structdemo}. 
These parameters include the average size (Col.\,2) and size dispersion (Col.\,3), the average elongation  (Col.\,4), and the average surface stellar density  (Col.\,5) of
all structures found in each density level. The total numbers of stars and total UV magnitudes of all structures in each level are also given (Cols.\,6 and 8 respectively). 
Table\,\ref{tab:structdemo} provides the fraction of stars at each significance level relative to the total young stellar sample (Col.\,7), and the corresponding  
stellar UV flux fraction relative to the total observed UV flux per detection level (Col.\,9). 


In general, almost all parameters given in Table\,\ref{tab:structdemo} show a dependence on the detection level\footnote{Among all parameters, only elongation, $\xi$, 
seems to be independent of the level of detection for the structures.}. The average size (and its dispersion), decreases with increasing density level, with a plateau 
and a small bump toward relatively larger sizes for levels between 6 and 12$\sigma$. The mean and median of the sizes reported in Col.\,2 of Table\,\ref{tab:structdemo}
equal  to 185\,pc and 138\,pc respectively. Both the total stellar number and total UV brightness 
show a systematic correlation with the detection significance level, with  larger and sparser stellar structures hosting higher stellar numbers and UV brightness. 
This agreement in the trends of these two parameters can be directly explained by the almost one-to-one correlation between their values, as can be derived from 
the data of Table\,\ref{tab:structdemo}. 

A systematic dependence on level also exists for the fraction $f_{\star}$ of stars included in every density level over the total observed number of blue stars. 
This fraction changes from 60\% within the 1$\sigma$ structures to $\sim$\,1\textperthousand\, at the highest density level (corresponding to one structure 
identified at 16$\sigma$ level). However, while the same dependence on significance level is found for the fraction of UV emission per level 
relative to that observed from all blue stars (Col.\,9 in Table\,\ref{tab:structdemo}), this trend is steeper than that for stellar fraction. 
In particular, while $\sim$\,60\% of the stars (i.e., the 1$\sigma$ structures) produces $\sim$\,70\% of UV, this correlation changes for structures 
with higher densities. For example, at 7$\sigma$ significance, where $\sim$\,7\% of the total young population resides, $\sim$\,15\% of the total UV brightness is produced.
At even higher levels only a few\,\textperthousand\, of the total young stellar population in NGC\,6503 emits few\,\% of the total UV light (see Table\,\ref{tab:structdemo}). 
This comparison suggests that  
compact structures, identified at higher density levels, encompass on average the UV-brightest stars in the galaxy. This result is elaborated more in
Sect.\,\ref{s:ACFevol}, where we discuss the clustering behavior of blue stars as a function of their brightness.



\begin{figure}
\centerline{\includegraphics[clip=true,width=0.475\textwidth]{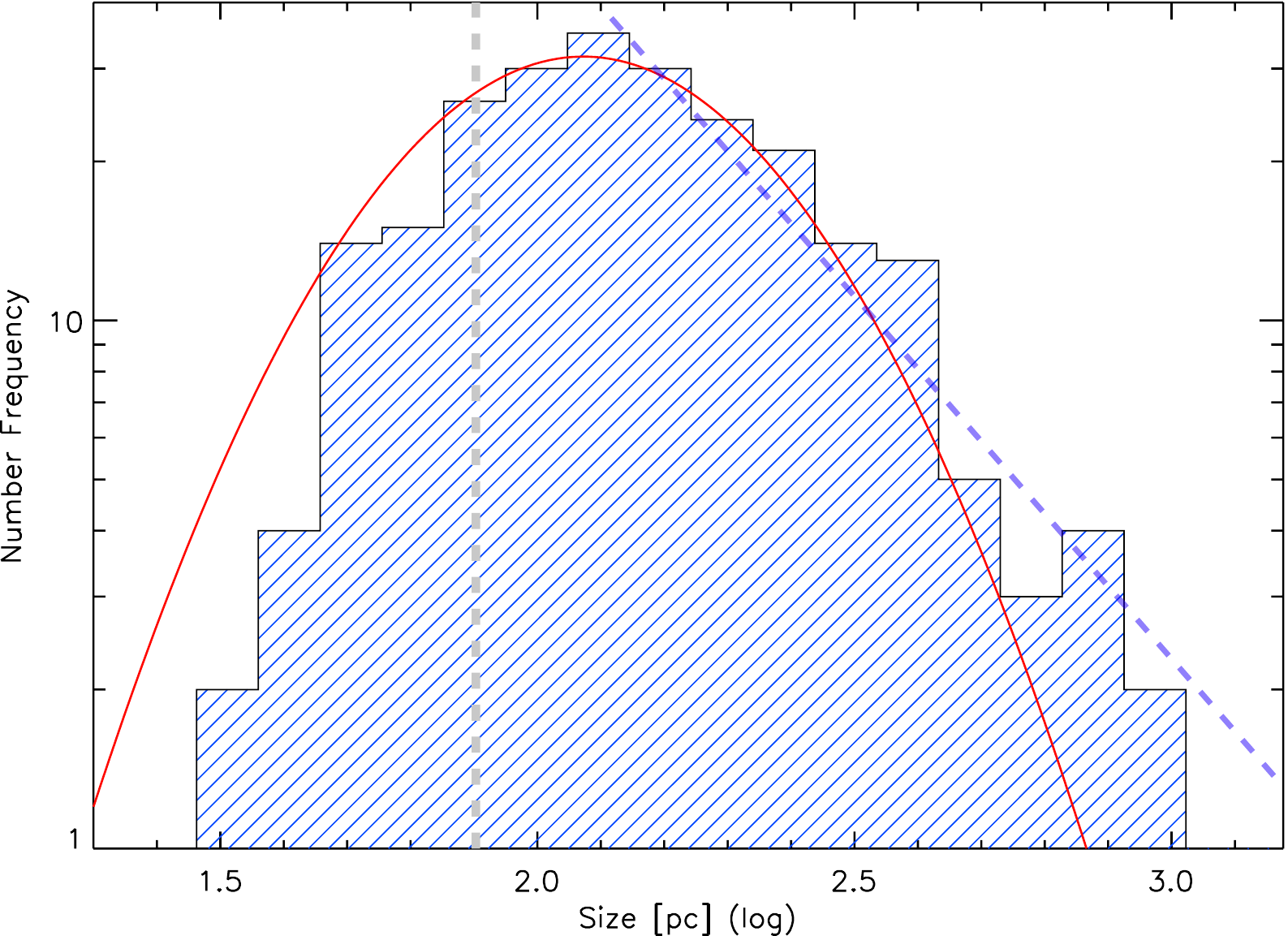}}
\caption{Histogram of the size distribution of all detected young stellar structures in NGC\,6503, 
with a logarithmic bin size of 0.075. Sizes are defined as $2 \times r_{\rm eff}$ and are 
given in pc. The distribution peaks at a size of $\sim$\,130\,pc. The best-fit Gaussian 
(red line) peaks at an average size of $\sim$\,120\,pc, close to the smoothing radius used for the 
structures identification of 80\,pc (indicated with the vertical grey dashed line). This fit shows 
that the right hand part of the size distribution, corresponding to the 
resolved structures, behaves almost like a power law (indicated by the blue dashed line).
\label{f:sizedist}}
\end{figure}

The size distribution of all detected structures constructed by binning them according to 
the logarithm of their dimensions is shown in Fig.~\ref{f:sizedist}. Dimensions are given in physical units (pc), and they
are derived from the effective radii of the structures. A functional fit of this histogram to a normal distribution, 
drawn with a red line, shows that the dimensions of the detected 
systems are clustered around an average of $\simeq 120$\,pc with a standard deviation of 
$\simeq 40$\,pc. This is an interesting result, considering that this size is comparable (but still to the upper limit) 
to the typical size for giant molecular clouds \citep[e.g.,][]{cox2000, tielens05}. It has been  
also pointed out earlier by various authors \citep[e.g.,][]{efremov87, ivanov96} that young stellar 
associations in different galaxies have dimensions that average at $\sim$\,80\,pc, comparable to 
the 40\% percentile of our distribution. Considering that the determination of structures dimensions is 
sensitive to the resolution of their detection, the question if this length-scale would represent a characteristic scale 
for star formation is still open \citep[e.g.,][]{bastian07, gouliermis11}. { Moreover, the detection 
resolution seems also to determine the peak of this distribution. By repeating 
the analysis after smoothing by smaller scales (between 70 and 40 pc), we found that indeed 
the peak moves at smaller values (from $\sim$\,120\,pc at 80\,pc 
resolution to $\sim$\,70\,pc at 40\,pc resolution).}


From this fit it is seen that the size distribution of our structures is not lognormal. There is 
a deficiency in the number of small-scale  structures at the left hand tail of the distribution, 
which falls below the Gaussian. { This deficiency is most probably due to incompleteness 
introduced by the smoothing, since it seems to be more important for the
distributions derived with larger smoothing kernels.} In addition, the right hand part of the distribution, 
for structures with sizes larger than $\sim$120\,pc, is more extended than the Gaussian and has a 
power-law shape. This is demonstrated with a power-law fit, shown with the blue dashed line, which 
indicates a scale-free behavior. { It is interesting to note that the power-law behavior of the right
part of the distribution seems to be independent of the smoothing kernel, having almost the same
exponent of $\sim -1.5$ for distributions produced after smoothing with smaller kernels.}

\begin{figure}
\centerline{\includegraphics[clip=true,width=0.475\textwidth]{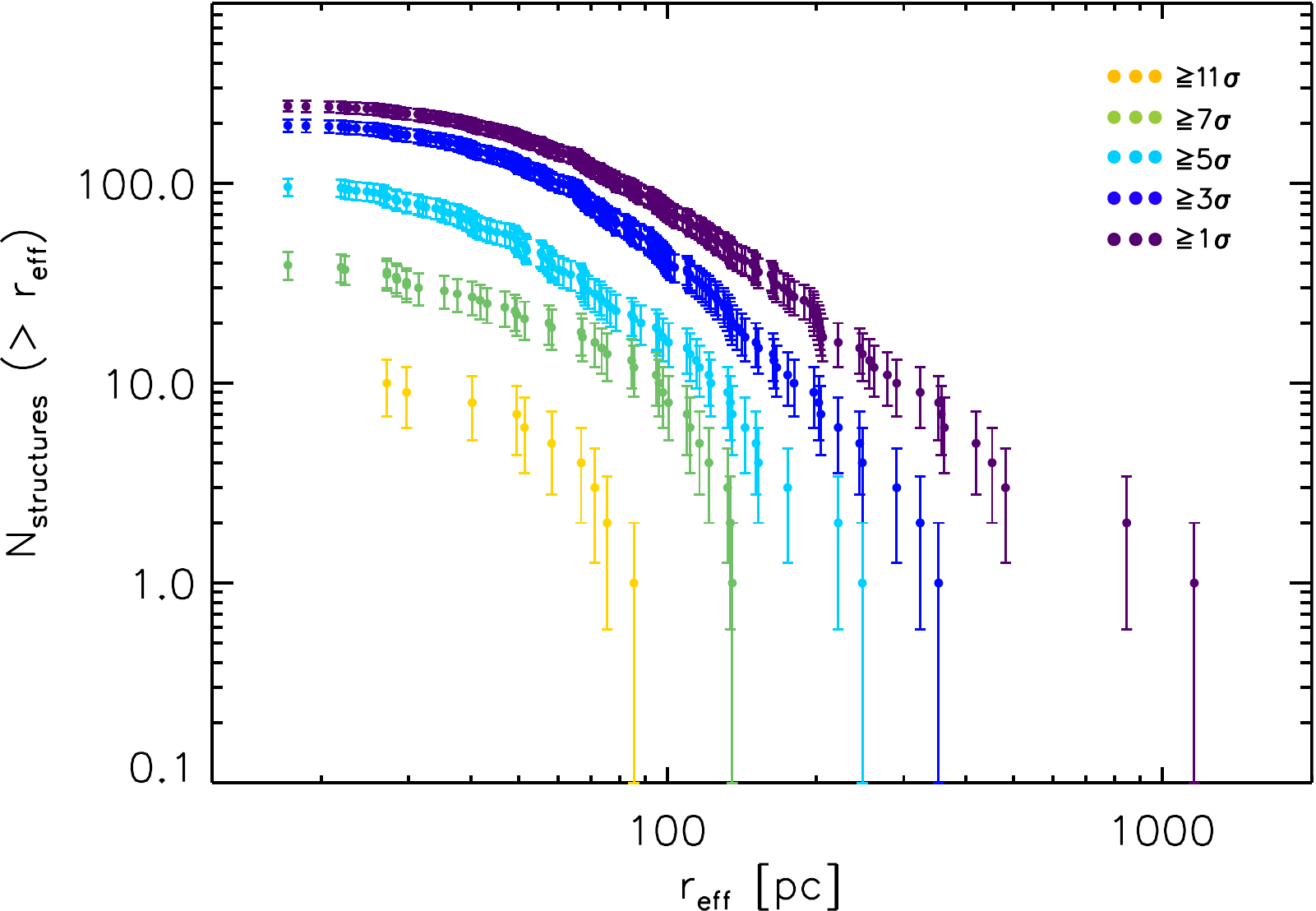}}
\caption{Radius cumulative distribution functions derived from the NGC\,6503 survey of star-forming structures. 
The results for five different detection density levels are shown. The distribution of structures found at levels $\geq$\,1$\sigma$ 
(purple symbols) corresponds to the total sample. This distribution and those for samples including low-density structures 
(blue symbols) show power-law tails at large radii. The distributions for samples constrained to high-density structures 
(green and yellow symbols) have shapes closer to lognormal.
\label{f:pdfrad}}
\end{figure}

This power-law behavior is further supported by the cumulative radius distribution 
of the star-forming structures found in different density thresholds (levels), shown in Fig.\ref{f:pdfrad}.  
The distributions of structures found only within five indicative detection significance levels are shown to 
avoid confusion. The cumulative distributions functions of radii for all structures  or those including low density 
structures (purple and blue symbols in Fig.\ref{f:pdfrad}) do not have lognormal shapes, at least in their entirety, 
with clear power-law tails at large radii. On the other hand, the 
distributions for detections at higher density thresholds (e.g., green and yellow symbols in Fig.\ref{f:pdfrad}) 
demonstrate a behavior closer to lognormal.

\subsection{Parameters correlations}\label{s:paramcol}

The correlation of physical parameters of star-forming structures provide insight to the processes and 
conditions that dictate their formation and evolution. If these systems do trace gas structure, then their 
fundamental properties should be tightly related to those of  GMCs,
and should have comparable correlations, which nevertheless should also be affected by 
the galactic dynamics and the transformation of gas to stars. In this section we explore the correlation of the 
structural parameters measured for the discovered stellar structures in order to identify the factors that 
determined those relations. The basic parameters considered are those directly measured from our data, namely 
the size of the structures (determined as two times their $r_{\rm eff}$), their surface stellar density,
$\mu$ (in stars\,pc$^{-2}$), the number of member stars, $N_\star$, and their total UV brightness, $m_{\rm 275}$, 
in magnitudes.  

We also  make a rough estimate of the volume stellar density of each structure, $\rho$ (in stars\,pc$^{-3}$), 
assuming spherical symmetry, as $\displaystyle \rho \propto \mu\, r_{\rm eff}^{-1}$. This estimation for elongated structures, 
which are typically the larger low-density structures, is expected to suffer from the lack of symmetry. However, we determine 
the volume density within the effective radius and not the maximum radius, which defines the largest 
aperture covered by the structure, and includes a large fraction of empty areas. Thus { we eliminate to a large extent the bias} in the volume density estimation due to asymmetries.

\subsubsection{Stellar Number Correlation with Radius}\label{s:r_nstars}

\begin{figure}
\centerline{\includegraphics[clip=true,width=0.475\textwidth]{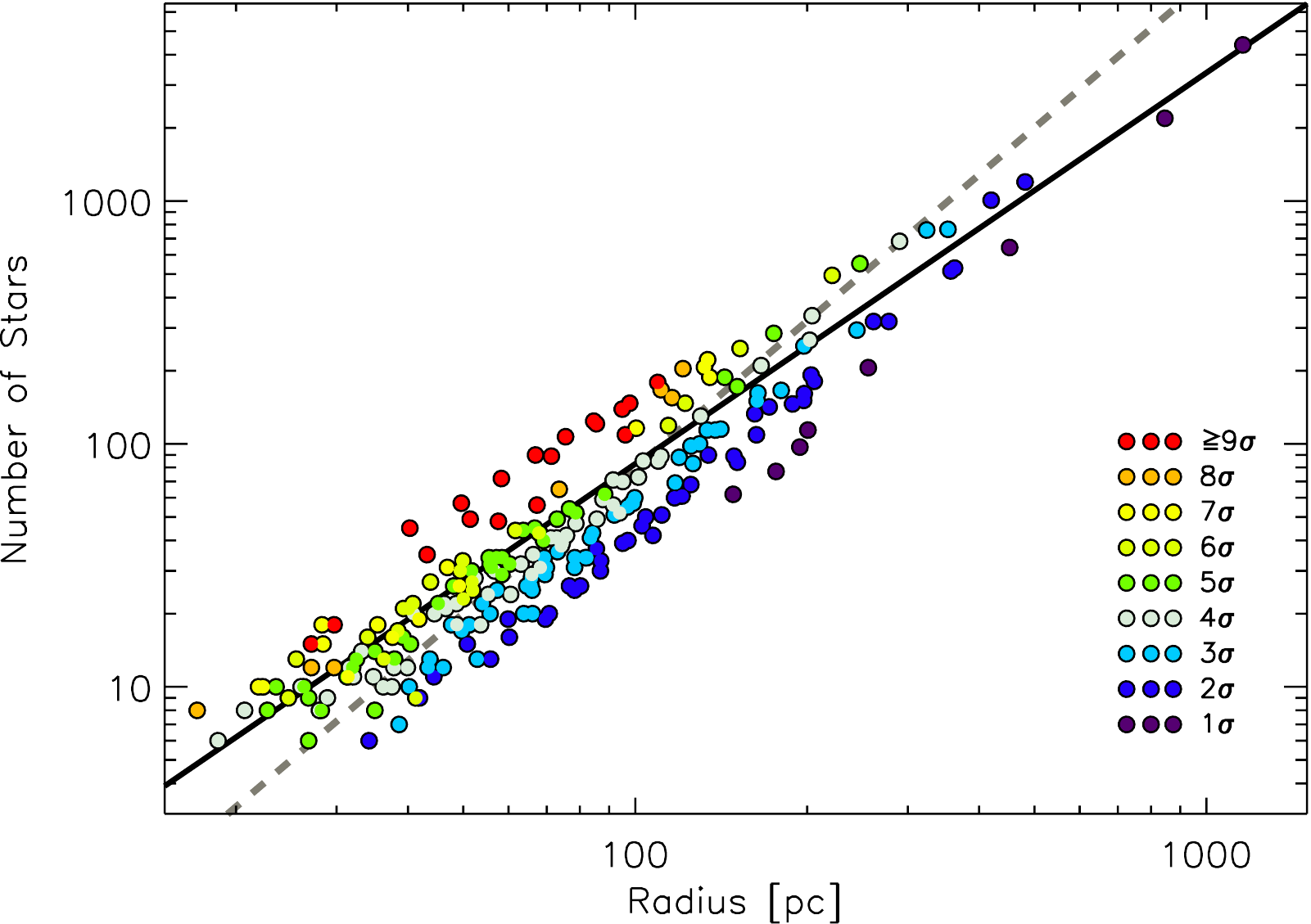}}
\caption{Correlation of number of stars with the radius, $r_{\rm eff}$, of the detected stellar structures. The different 
colored points represent structures identified at different density levels, indicated by the legend.
The black solid line is the best fit through all the data, with a slope of 1.6. This fit for the various significance 
structures varies in agreement with previous investigations. This power-law behavior is expected from a fractal 
stellar distributions. A slope of 2, corresponding to constant stellar surface density, is shown with the grey dashed 
line to indicate how a uniform distribution appears in this correlation. 
\label{f:sizenstars}}
\end{figure}

To measure molecular clouds structure \cite{kauffmann10a, kauffmann10b} established the correlation between mass and radius 
for clouds in the solar neighborhood. Using the survey of stellar complexes of the present work, 
we can test such a relation for large star-forming structures. As a first-order approximation, the number of stars in a group is directly 
proportional to its mass, assuming all groups are of a similar age and as long as sampling effects of the mass function are not significant. 
In Fig.\,\ref{f:sizenstars} we show the measured radii, $r_{\rm eff}$, versus the number of stars within each identified structure. The solid line 
in the plot represents the least-square linear fit to the logarithms of the data. { While there is no significant scatter} in the whole sample, 
the relation for systems identified at lower density thresholds is different than that for those found at higher densities with indexes  
varying between about 1.5 and 2. This trend is consistent to that found by \cite{gouliermis03} for 494 stellar associations and open clusters in the Large 
Magellanic Cloud.

A single power-law mass-size relation with a fractional index is expected from a fractal distribution of 
young stars \citep[in agreement with, e.g.,][]{elmefalga96}. A uniform two-dimensional distribution of stars
would correspond to a constant surface stellar density, and therefore the number of stars 
(or mass) would be proportional to the radius (or size) as $\propto \pi r^{2}$. Thus, a uniform stellar distribution 
would also produce a power-law behavior but with an index 2 due to the dependence of surface to radius. 
This relation is shown with a dashed line in Fig.\,\ref{f:sizenstars}. 
Concerning structures in our sample, the correlation between number of stars and radius for the low-density structures (up to 3$\sigma$)  
has slopes very close to 2, indicating that these systems have constant surface densities. On the other hand, the high-density systems show correlations different 
to a uniform distribution with slopes between 1.5 and 1.8 (all slopes have about 1\% uncertainty), consistent with self-similar stellar distributions. Such a dependence of the 
mass-radius relation to the detection threshold is also found in M\,33 by \citet[][who used a different technique]{bastian07}, summarizing the differences in the trends 
of properties for different types of stellar systems. We explore these differences in the following section.


\begin{figure}
\centerline{\includegraphics[clip=true,width=0.475\textwidth]{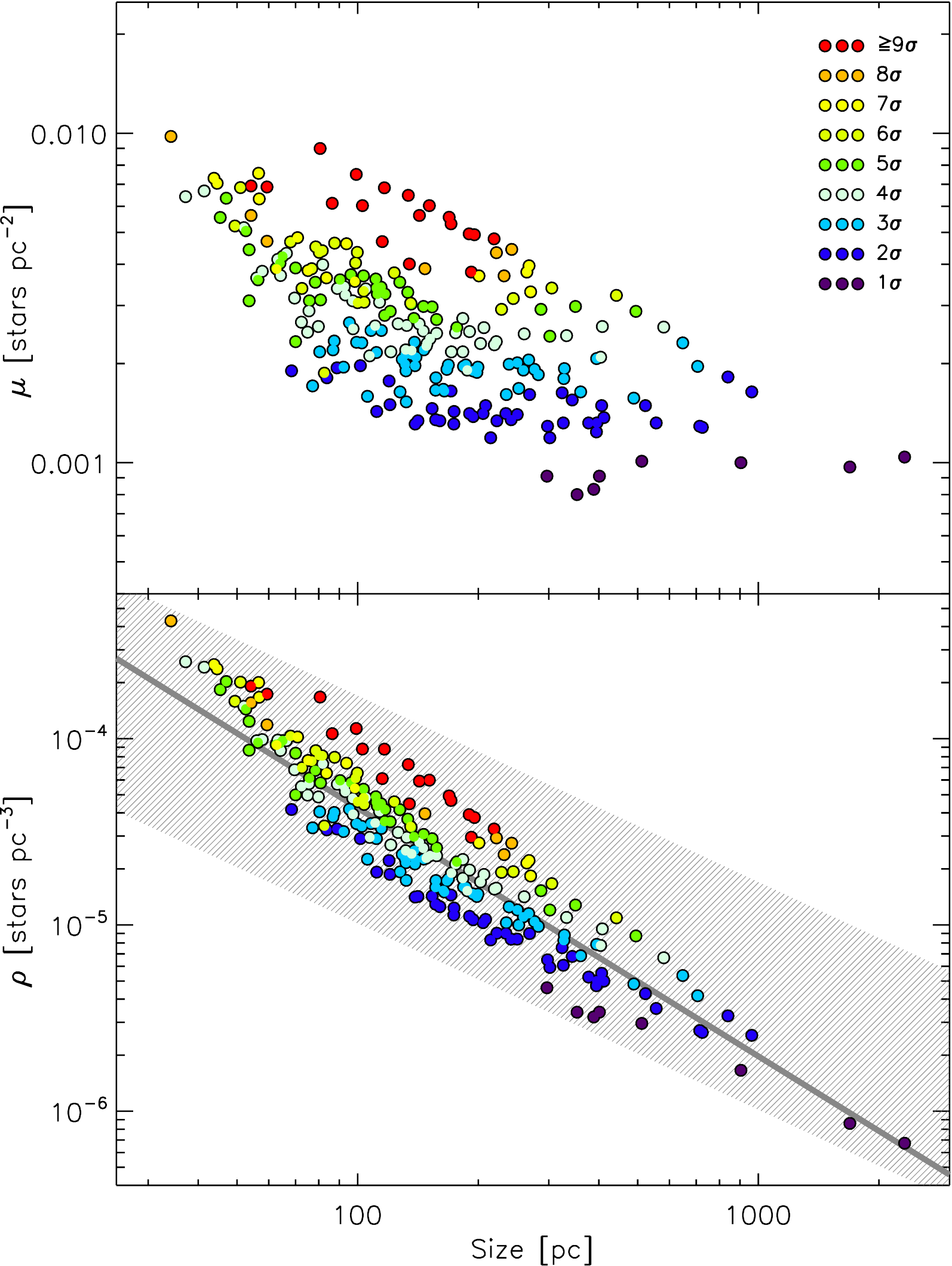}}
\caption{Surface stellar density (top) and volume stellar density (bottom) correlation with the size of 
the detected stellar structures. In the top panel it is shown that the surface density of low-density (large) 
structures, in particular those detected at levels lower than $\sim$\,5$\sigma$, is almost constant, in contrast to 
the high-density systems, where a correlation of surface density and size is more apparent. On the other hand, 
volume density, plotted in the bottom panel, shows a clear correlation with size for all structures, with 
different exponents for structures found at different density levels. The shaded area represents 
Larson's 3rd relation, which correlates the sizes of molecular clouds to their volume densities. 
The correlation found for our stellar structures seems to comply with this relation, although it has a trend 
to be on average steeper (fit is shown with the thick line). The correlation for systems detected at 
levels lower than $\sim$\,5$\sigma$ is more compatible to Larson's relation than those for systems found 
at levels $\gsim 5\sigma$, which are steeper (see also Fig.\,\ref{f:levvspowidx}). 
\label{f:sizedenscorr}}
\end{figure}

\begin{figure}
\centerline{\includegraphics[clip=true,width=0.475\textwidth]{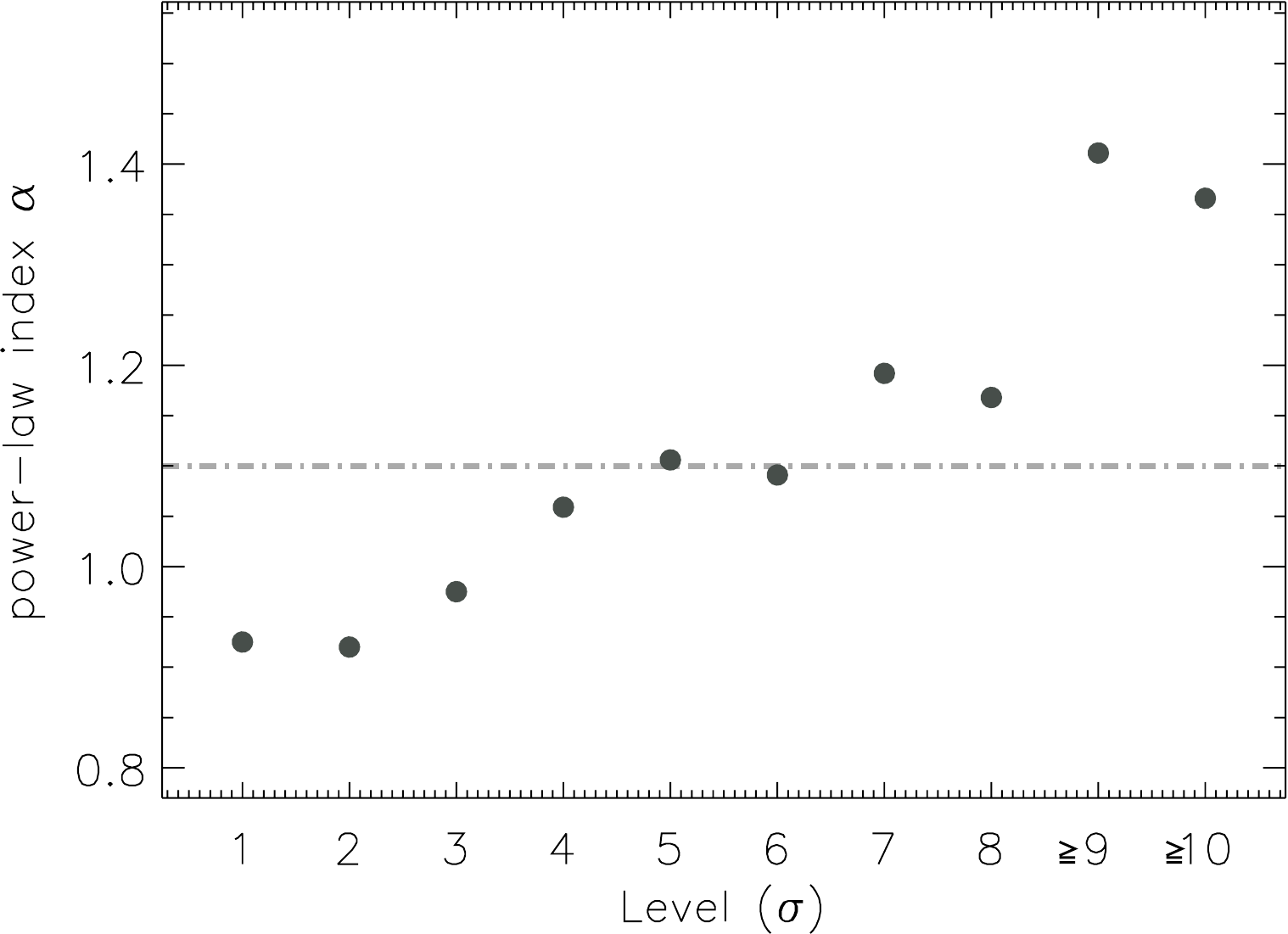}}
\caption{Relation between the index $\alpha$ in the relation $\rho\propto L^{-\alpha}$ for the detected stellar structures 
and the corresponding detection density level (in $\sigma$). This plot shows a dependency of the correlation between
volume stellar density and size of the structures to their detection limits. The size of large, low-density structures 
(found at levels up to $\sim$\,5$\sigma$), seems be correlated to their corresponding volume density in a fashion
comparable to or flatter than Larson's third relation (indicated by the horizontal dashed dotted line), while high-density, 
smaller structures have clearly steeper correlation between their sizes and volume densities.
\label{f:levvspowidx}}
\end{figure}

\subsubsection{Surface and Volume Stellar Density Correlations with Size}\label{s:rho_sz}

Correlations between the surface stellar density $\mu$  and volume stellar density $\rho$  
of the young stellar structures to their sizes $L$ (in pc) are shown in Fig.\,\ref{f:sizedenscorr}. Points
are colored according to the detection surface density level  (in $\sigma$) of the corresponding structures.
Both plots show an overall dependence of stellar density on size, with that for the volume density being more 
prominent. However, if structures are grouped according to their density detection levels, at the top panel of  
Fig.\,\ref{f:sizedenscorr} it can be seen that there is no clear dependence of surface density on size for low-density 
structures, in particular those identified at levels $<$\,5$\sigma$, showing almost flat distributions of points. This is not 
the case for the high-density (more compact and smaller) structures. 


The volume density-size correlation, shown at the bottom panel of  Fig.\,\ref{f:sizedenscorr}, can be represented for all points 
by a power law of the form $\rho\propto L^{-\alpha}$ 
with index $\alpha= -1.33$, established with a linear fit to the data (grey thick line in the plot). 
However, as can be seen in the plot, this slope differs  
for structures identified at different density detection levels, with low-density (and larger sizes) structures following 
a more shallow relation than the high-density structures. We demonstrate the dependency 
of the exponent $\alpha$ for our stellar structures to their detection density level in Fig.\,\ref{f:levvspowidx}, 
where we plot $\alpha$, as found for every group of structures, as a function of the corresponding density threshold 
(in $\sigma$).



The correlation of volume densities and sizes we establish for the detected stellar structures 
can be directly compared to that determined by \citet{larson81} for molecular clouds, and confirmed 
by other authors \citep[e.g.,][]{myers83, solomon87, heyer09}\footnote{It should be noted, though, 
that while our structures for detection levels $\leq$\,6$\sigma$ have sizes larger than $\sim$\,100\,pc, 
most of the molecular clouds in the Milky Way have sizes of that order or less.}. The so-called 
Larson's third relation exhibits the tendency of the mean cloud volume density, $\rho$ 
(in cm$^{-3}$) to scale inversely  with the cloud size, $\rho\propto L^{-1.1}$. 
Larson's exponent of $-1.1$ is indicated by a horizontal dashed-dotted line in Fig.\,\ref{f:levvspowidx}.
This exponent coincides with $\alpha$ determined in our correlation for structures at  density 
levels between $\sim$\,5$\sigma$ and 6$\sigma$. 


Larson's relation implies that clouds have approximately constant column densities, 
i.e., the same significance levels relative to a fixed background, reflecting the identification of 
molecular clouds as constant-density objects  \citep{solomon87}. In the case of our stellar 
structures, as seen in Fig.\,\ref{f:sizedenscorr} (top panel) this does not apply, since 
the identification of these structures is made at various density levels. Only structures detected 
at low density levels ($<$\,5$\sigma$) have surface stellar densities almost independent of their size, 
and  $\rho(L)$ relations comparable to Larson's relation ($\alpha$ values below the 1.1 limit in 
Fig.\,\ref{f:levvspowidx}). In contrast, high-density structures have steeper $\rho(L)$ relations, 
because their $\mu(L)$ relations are not flat. These steeper correlations are the result of the 
higher densities of structures that tend to be smaller, and reflect the steeper density profiles 
of these systems (as $\propto r^{-\alpha}$ for spherical systems) in relation to those of low density.


%
%

\begin{figure}
\centerline{\includegraphics[clip=true,width=0.475\textwidth]{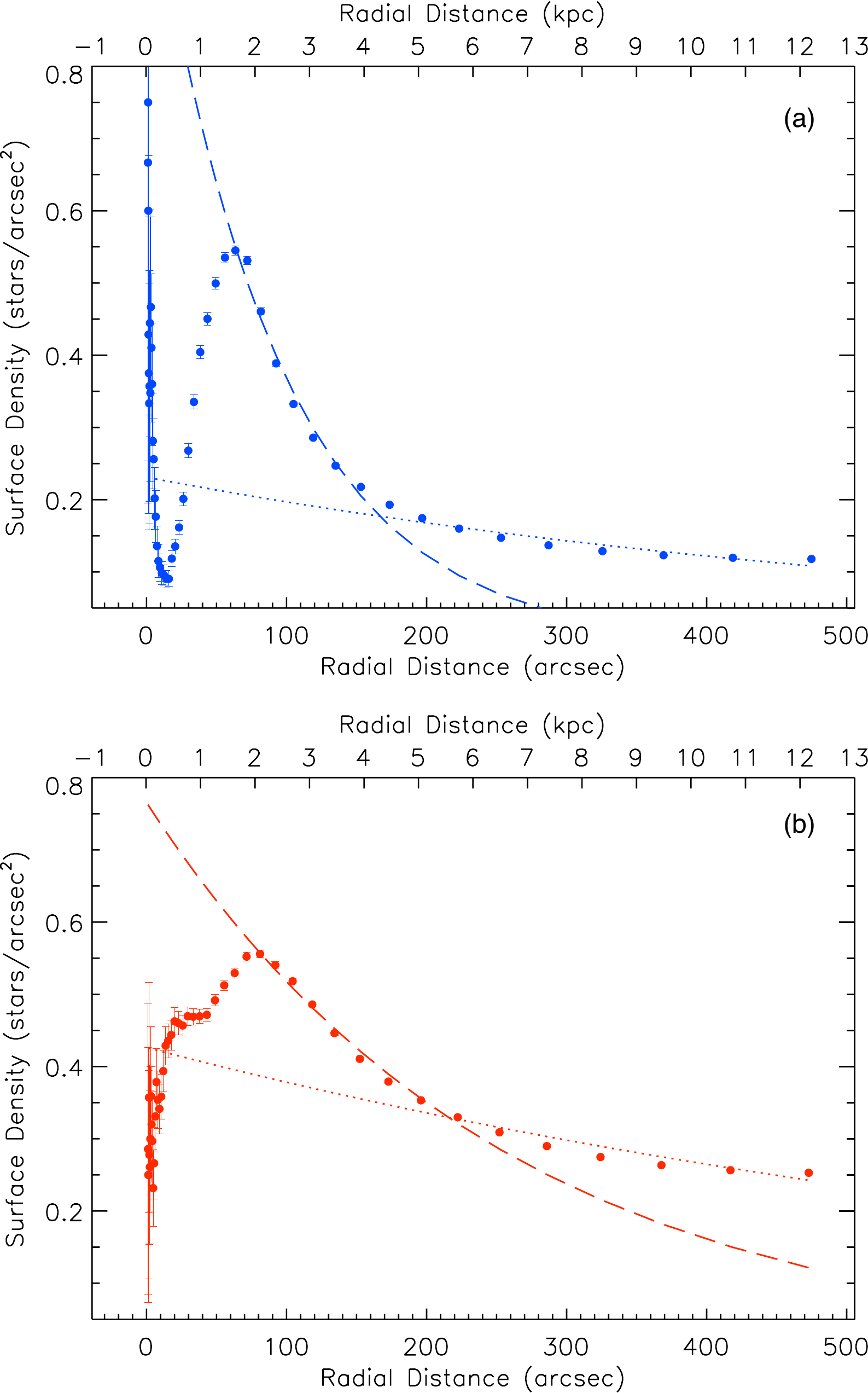}}
\caption{(a) Cumulative radial stellar density profile of the blue stellar population in NGC\,6503. This
profile exhibits the ring-shaped distribution of the youngest populations, highlighting the very low 
abundance of young stars in the inner parts of the galaxy. In this profile the star-forming ring peaks at a radial 
distance $\sim$\, 72\arcsec$_{-8}^{+10}$, corresponding to $\sim$\,1.85$_{-0.21}^{+0.26}$\,kpc. (b) The corresponding 
profile of the red population shows a different behavior, with a much lower deficiency of stars in the inner parts, and
a peak at somewhat larger radial distance of $\sim$\, 81\arcsec$_{-10}^{+11}$ ($\sim$\,2.09$_{-0.26}^{+0.28}$\,kpc). 
Exponential disk fits of the profiles are shown with dashed lines. These fits, equivalent to those on the surface brightness 
profiles previously presented \citep{bottema89, freeland10}, do not fit the outskirts of the observed distributions, which are fitted 
with different exponential profiles (shown with dotted lines).
\label{f:radprof}}
\end{figure}

\section{The Global clustering of young stars in NGC\,6503}\label{s:hierarchy}

\subsection{Stellar Surface Density Profiles}\label{s:ssdp}

We construct the radial stellar surface density profile across the observed field-of-view of the blue stellar population of NGC\,6503, by
counting the stars in concentric annuli around the rotational center of the galaxy (defined in Sect.\,\ref{s:deproject}). 
The annuli are determined 
in logarithmically increasing radial distances, so that the width of each annulus increases with its radial distance. The cumulative 
radial stellar surface density profile of  the blue stellar population of NGC\,6503 is shown in Fig.\,\ref{f:radprof} (top panel).
In this figure the profile of the red population, as this stellar sample is determined in Sect.\,\ref{sect:selection},
is also shown for comparison (bottom panel).

From these profiles it is seen that blue and red stars follow quite different radial distributions around the center of the galaxy.
Specifically, the blue stars show a very strong peak at the center, where the LINER is located. The surface density drops 
rapidly within a very short radial distance ($\sim$\,14\arcsec\,$\sim$\,0.35\,kpc) and then rises sharply, reaching a peak at a 
distance of $\sim$\,1.9\,kpc (Fig.\,\ref{f:radprof}, top). The distribution of the red 
stars shows also central densities lower than those at larger radii, but their deficiency in the inner part 
of the galaxy is not as severe as for the blue stars. Nevertheless, the surface density of red stars rises slowly outwards and peaks at a 
distance of $\sim$\,2.1\,kpc, a distance comparable to that where the blue stellar distribution peaks. It drops again at larger distances, 
but far more slowly than that for the blue stars (Fig.\,\ref{f:radprof}). 

We evaluated the deficiency of stars in the galaxy center in terms of incompleteness in our photometry due to crowding.
Our artificial star experiments showed that { the lack of blue stars in the the circumnuclear region} is not due to photometric incompleteness.
The number of UV sources is apparently very small and thus the blue stellar sample is flux- rather than crowded-limited 
even at the center of NGC\,6503. The situation is different in the optical filters, as for example in the F555W 
filter, which is found to be clearly more incomplete in the center than in the periphery. As a consequence, { the circumnuclear decreases} seen in the stellar profiles of Fig.\,\ref{f:radprof} are probably due to different reasons. While the central lack of blue sources 
is due to the real absence of a significant (and dense) young population in the center, the central deficiency of red stars 
is due (at least in part) to incompleteness.

Both surface density profiles shown in Fig.\,\ref{f:radprof}, and in particular that of the blue young population, track the 
inner star-forming ring of the galaxy with a peak at $\sim$\,2\,kpc. This is in line with the previously published radial luminosity profile
of NGC\,6503, which is proposed by \cite{bottema89} as a 
typical example of Type II \citep{freeman70}, i.e., of the form $I(R) < I_{0}e^{-\alpha R}$ for a 
radial distance interval not far from its center. At larger radial distances both our blue and red profiles drop in an 
exponential fashion, as shown by the fitted exponential disk models plotted with dashed lines on the observed 
profiles of Fig.\,\ref{f:radprof}. These exponential disk models agree with those describing the smooth surface brightness profile of the 
galaxy in various wavelengths for radial distances up to about 3\,kpc \citep{freeland10}. 
However, in order to describe the distribution of stars 
at the outskirts of our stellar surface density profiles for radial distances \gsim\,4-5\,kpc,  fits to additional exponential disks were considered.  
These fits are plotted in the profiles of Fig.\,\ref{f:radprof} with dotted lines. 



A Type\,II profile produced by a stellar deficit in the inner parts of the galaxy may be caused by (1) dust extinction, 
(2) an inner truncated disk, or (3) a ring of bright stars. An exponential disk and a truncated light profile due to dust extinction are proposed for 
NGC\,6503 by \cite{bottema97}, but a statistical study on several galaxies showed that the possibility of dust extinction causing Type II profiles 
is inconclusive \citep{macarthur03}. Concerning the other two possibilities, a truncated disk in NGC\,6503 implies that there is a genuine mass 
deficit in the inner region of the galaxy, and a stellar ring relates to bar formation. The dynamics of NGC\,6503 was modeled by \cite{puglielli10} with a Bayesian 
inference by assuming a \cite{kormendy77} truncated disk. These authors found that the disk of NGC\,6503 is indeed equally well-fitted by either mechanisms, i.e., 
an inner-truncated profile, or a ring formation by a bar. 


The inner drop of stellar density seen in our profiles also agrees with a sharp decrease in the velocity dispersion ($\sigma$-drop) for radii smaller 
that $\sim$\,10\arcsec ($\sim$\,0.25\,kpc) observed by \cite{bottema89} with a minimum at a distance remarkably similar to that of our blue stellar 
surface density profile\footnote{The velocity dispersion drop in NGC\,6503 is also discussed by \cite{bottema97} and \cite{puglielli10}}. Modeling 
of the dynamics of the galaxy, however, could not provide a reliable theoretical description of the inner drop in dispersion \citep{bottema97}.
In their study \cite{puglielli10} found that the mass-luminosity ratio of NGC\,6503 pseudobulge is lower than that in the disk, suggesting the 
presence of a dynamically-cold star-forming component that is probably responsible for the velocity dispersion drop, in agreement with previous 
studies \citep{wozniak03, comeron08}.  The existence of an inner nuclear bar could also produce the $\sigma$-drop, as has been reported in the 
literature for some cases \citep{delorenzo-caceres08}. Indeed, while the ring of NGC\,6503 is not considered to have the 
classical aspect of dynamically induced resonance rings \citep{mazzuca08}, it is likely an {\em inner ring and not a nuclear ring}, 
possibly caused by a strong end-on bar which is embedded inside it \citep{knapen06, freeland10}. 




\subsection{Hierarchical Structure in Stellar Clustering}\label{s:ACF}

As discussed in Sect.\,\ref{s:structident}, young stellar structures in NGC\,6503 are assembled in a hierarchical fashion,
in the sense that there are condensed structures belonging to larger looser ones, which themselves are part of even larger 
low-density stellar systems. This behavior, demonstrated visually in the dendrogram of Fig.\,\ref{f:dendrogram}, suggests that 
the morphology of { young stellar cluster assembling} across the whole galaxy is self-similar. In this section we quantify the young stellar clustering 
behavior with the use of the two-point correlation function and we determine the time-scale within which stellar clustering 
sustains its behavior.


The spatial distribution of stars can be quantified with the construction of the two-point correlation or autocorrelation function (ACF), 
which is a measure of the degree of clustering in the spatial distribution, $\xi(r)$, of a sample of sources \citep{baugh00}. This method, 
introduced by \cite{peebles80} in cosmology, has been successfully used 
for characterizing the stellar clustering behavior in star-forming regions in the Milky Way \citep[e.g.][]{gomez93, larson95}, as well as 
that of stellar populations and star clusters in remote galaxies \citep[e.g.,][]{bastian09lmc, scheepmaker09}. In this study we apply the method to the resolved 
stellar population of a whole galaxy, following the prescription by \cite{gouliermis14}. The innovation of our treatment lies 
on the use of the {\em de-projected} positions of the stars, eliminating the effect of projection on the measures of stellar pair-separations. 
The constructed ACF is, thus, the true two-dimensional ACF of the galaxy, as if it was 
observed face-on. We determine the ACF of the stars from their de-projected coordinates, as: 
\begin{eqnarray} \label{eq:autocorrelation}
1+\xi (r) & =  & \frac{1}{\bar{n}N}\sum_{i=1}^{N} n_{i}(r),
\end{eqnarray} 
where $N$ is the total number of stars, $n_{i} (r)$ is the number density of stars 
found in an aperture of radius $r$ centered on star $i$,  and $\bar{n}$ is the
average stellar number density. Uncertainties of this function are given by:
\begin{eqnarray}\label{eq:acferr}
\delta(1+\xi (r)) & = & \sqrt{N} \cdot\left(\frac{1}{2}\sum_{i=1}^{N} n_{\mathrm{p}}(r)\right)^{-1/2}, 
\end{eqnarray}
where $n_{\mathrm{p}}(r)$ is the number of pairs with the
central star $i$ of the current aperture, and the factor $1/2$ accounts for not counting every pair twice.
The function $1+\xi (r)$, is defined so that $\bar{n}[1+\xi(r)] d^2r$ is the probability of
finding a neighboring star in an area of radius $r$ from a random 
star in the sample. Therefore, for a random stellar distribution $1+\xi(r) = 1$, 
while  a truly clustered sample should have  $1+\xi(r) > 1$. 

\begin{figure}
\centering
\includegraphics[width=\columnwidth]{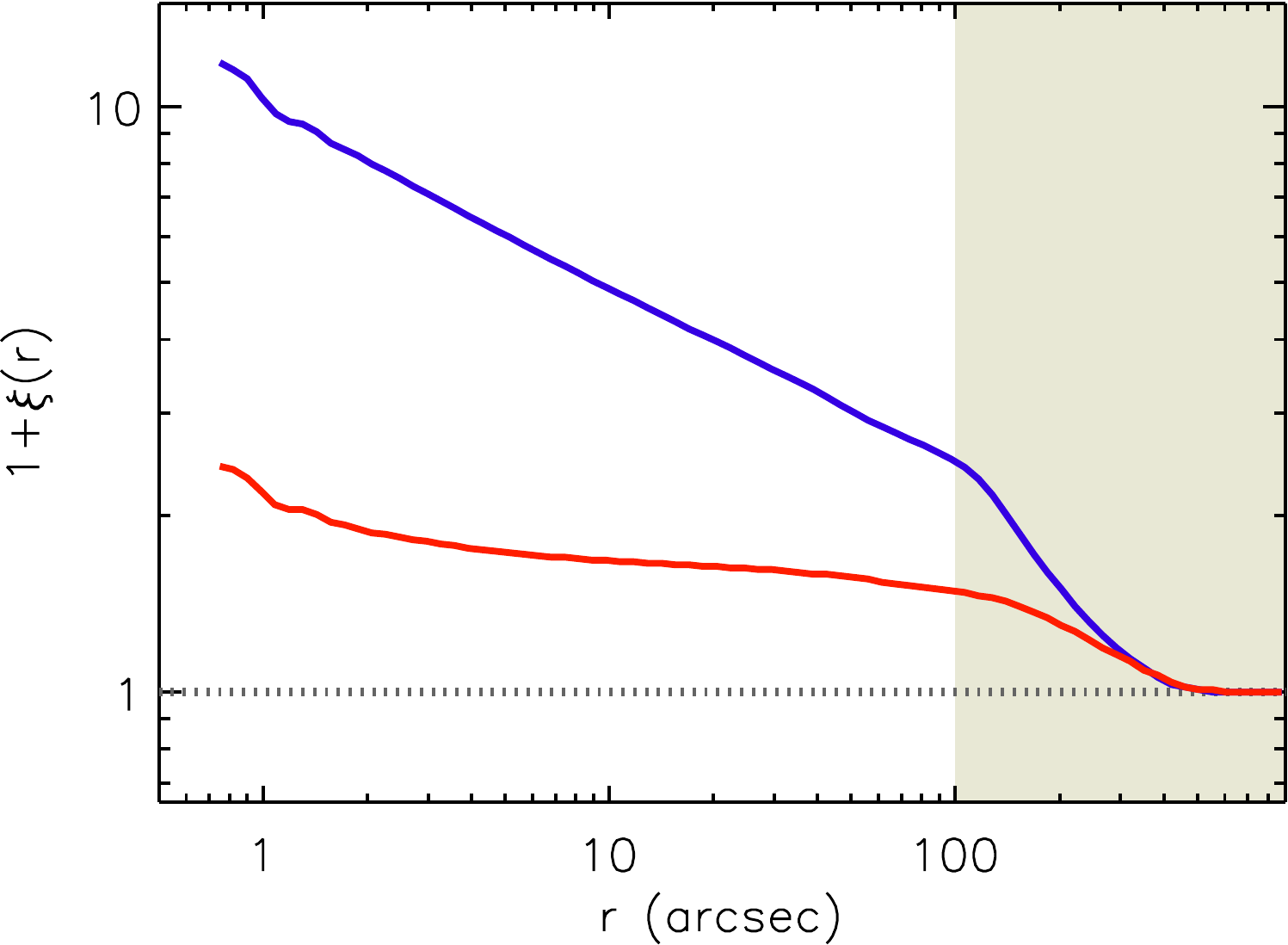} 
\caption{The two-point correlation (autocorrelation) function constructed for the two stellar samples of interest, 
revealed with LEGUS photometry in NGC\,6503. The ACF of the young blue stellar population detected in F275W and F336W
filters is shown with the blue line, while that of the older red stars selected in filters F555W and F814W is drawn 
with the red line. The grey flat dotted line represents the autocorrelation function of a uniformly distributed stellar population. From this plot 
it is shown that the young blue population follows a clustering behavior, which is quite different than that of the evolved red stars, being 
{\em more clustered} than the latter. The ACF of the red stars, being almost flat,  demonstrates that these stars are quite {\em more distributed}
than the blue. The shaded area designates the length-scale, where the ACF calculation is unreliable -- and thus
not considered in our analysis -- due to the edge-effect introduced by the limited field-of-view.
\label{fig:aACF_plots}
}
\end{figure}

In a two-dimensional self-similar distribution the total number of  stars $N$  within an aperture of radius $r$ increases 
as $N\propto r^{D_{2}}$, where $D_{2}$ is the fractal dimension of the 
distribution \citep{mandelbrot83}. Fractal distributions have a power-law dependency of the ACF with 
radius of the form $1+\xi (r) \propto r^{\eta}$ \citep[e.g.,][]{larson95}, which yields from Eq.\,(\ref{eq:autocorrelation}) 
$N\propto r^{\eta}\cdot r^{2}  = r^{\eta+2}$. The exponent $\eta$ is thus related  to the fractal dimension as $D_{2} = \eta +2$.
For the derivation of the three-dimensional fractal dimension $D_{3}$ of a distribution, dedicated
simulations are shown to be required \citep{gouliermis14}. The ACF is qualitatively identical method 
to a power spectra analysis. Both statistics are used in studies of cosmological large-scale structure 
\citep[e.g.,][]{szapudi05} and of density structure in the turbulent interstellar matter \citep[e.g.,][]{federrath09}.

In Fig.\,\ref{fig:aACF_plots} we show the ACF of the blue stellar sample observed in NGC\,6503 (blue  line).
For comparison we also show in red the ACF of the red stellar sample.  
Both functions were possible to be constructed for separations down to 0.75 seconds of arc, corresponding to 
physical scales of $\sim$\,20\,pc. For smaller scales, there is an ``anti-correlation'' of the ACF with separation, 
which drops rapidly towards the smallest separations. Apparently, photometric confusion dominates at scales smaller
than 20\,pc, a limit which, thus, sets  the resolution of our analysis with these data\footnote{Spatial scales $< 20$\,pc, 
corresponding to the cluster-formation regime, will be addressed with the star clusters census of NGC\,6503, which 
is currently under construction (Adamo et al., in prep).}. 
The ACF, being determined through the pair-separations between all stars, can be calculated  
within a specific maximum length-scale limit,
determined by the length of the unavoidably finite observed field-of-view.
In practice, the continuously larger stellar pair-separations calculated for each star in the sample eventually 
reach the edge of the observed field, beyond which there are no stars available for the calculation.
If the ACF determination is not corrected for this edge-effect, the calculation will be incorrect at large scales with 
its value dropping to the unrealistic $1+\xi(r) < 1$. 

We correct our calculations by masking the pair-separations 
measurements within the borders of the observed field-of-view, as prescribed by \cite{gouliermis14}. 
This treatment corrects the stellar densities of Eq.\,(\ref{eq:autocorrelation}) by using the true area surface 
constrained for the larger length-scales by the masking. This correction allows the ACF to fall smoothly 
to the limiting value of 1 at scales comparable to the edge of the observed field as shown in the plots 
Fig.\,\ref{fig:aACF_plots}. In effect, the sharp drop to the limiting value of 1 sets the maximum length-scale  
within which the ACF is reliable. We denote in Fig.\,\ref{fig:aACF_plots} with the greyed area the remaining unreliable 
length-scale range, which we do not take into account in our further analysis. It is interesting to note that the maximum 
scale, where a trustworthy ACF is calculated is $\simeq$\,100\arcsec\,($\simeq$\,2.6\,kpc), far shorter 
than the complete extent of the de-projected field-of-view.

We determine the exponent $\eta$ of the ACF through a linear regression on the log-log plots of 
Fig.\,\ref{fig:aACF_plots}  by applying a Levenberg--Marquard nonlinear least square minimization 
fit \citep{levenberg44, marquardt63}.  In this figure it is seen that both the blue and red stellar samples in 
NGC\,6503 have a single power-law stellar separations dependency of their ACF. The exponent 
$\eta$, however, is quite different between the two samples. The ACF of the young blue sample 
has a slope $\eta \simeq -0.30$, corresponding to a fractal dimension $D_{2} = 1.7$. On the other 
hand, the ACF of the old red stellar sample is almost flat with $\eta \simeq -0.05$, i.e., $D_{2} = 
1.95$ (both slopes found with very small fitting errors). 

There are two conclusions, connected to 
each other, one can derive from these results. (1) The steep monotonic ACF of the blue stars, 
corresponding to a fractal dimension much smaller than the geometrical dimension of 2, suggests 
a fractal, i.e., self-similar distribution of young stars across NGC\,6503. On the other hand, the flatter 
ACF for the red stars, with a fractal dimension very close to the geometric dimension, clearly implies 
a distribution for the old population, which is very well spread, almost equivalent to a uniform (random)
spatial distribution\footnote{The derived value for the ACF  of the red population serves as a confirmation 
of the capability of the method to distinguish a randomly dispersed sample of stars from any other distribution 
(fractal or not), by its flat ACF, which corresponds to a fractal dimension comparable to its geometrical one 
\citep[see, e.g.,][for comparisons with the ACF of synthetic fractal and centrally condensed 
distributions]{gouliermis14}.}. (2) The absolute values for the ACF of the young stars, which are found systematically 
higher than those for the old stars at the same separation length, indicate the more ``clumpy'' clustering 
behavior of the former sample in comparison to the latter, with the young stars being systematically {\em 
more clustered} than the old\footnote{The clumpy clustering of the young stars is tightly connected to their 
self-similar distribution; The more fractal a distribution is, i.e., with lower $D_2$ values, the more prominent 
sub-clustering behavior its stars have, i.e., higher ACF values \citep{gouliermis14}.}.

The steep ACF of the young stellar population in NGC\,6503 demonstrates that these stars are
hierarchically distributed across the whole measurable extend of the galaxy, up to scales $\sim$\,2.5\,kpc.
This implies that the hierarchical stellar structures identified in Sect.\,\ref{s:method}, are themselves 
part of an extended hierarchical distribution. Power-laws similar to that found here for the ACF of 
young stars in NGC\,6503 are derived from power spectra of interstellar gas over a large range of 
environments \citep[e.g.][]{elmegreen01}, demonstrating the hierarchical morphology in the 
gas structure with a typical fractal dimension of $D_{2}\simeq 1.5$ \citep[][]{elmegreen06}. 

\begin{figure}
\centering
\includegraphics[width=\columnwidth]{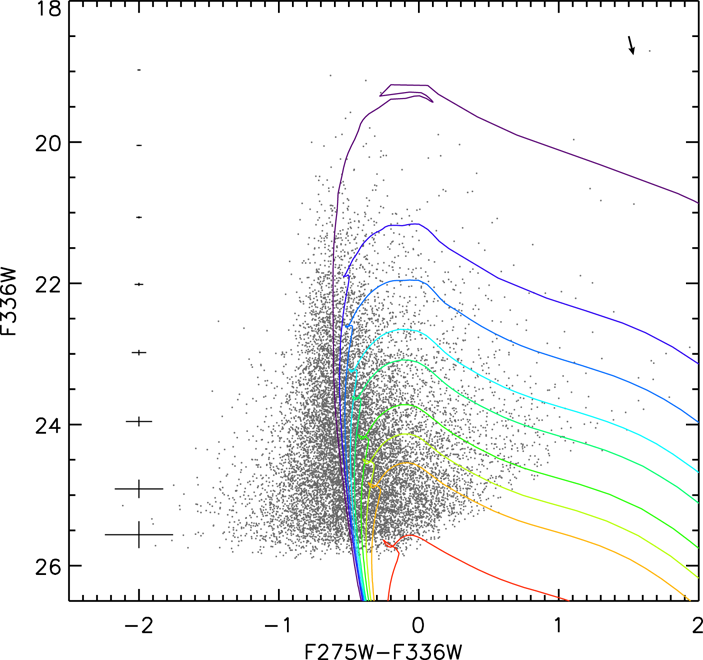} 
\caption{Total F275W, F336W CMD of the blue stars detected in NGC\,6503 with our photometry, 
with indicative isochrones from the Padova family of evolutionary models overlaid. Indicative isochrone models
corresponding to stellar ages between 4\,Myr (purple line) and 110\,Myr (red line) are shown.
The crosses on the left represent typical uncertainties in brightness and color in our photometry.
Isochrones are chosen for solar metallicity, and they are corrected for distance (with a distance modulus 28.6\,mag) 
and foreground extinction, represented by the arrow on the top-right of the diagram, according to the reddening law 
by \citet{fitzpatrick99}. Only the main-sequence and sub-giant branch are shown for each of the models for clarity. 
This plot demonstrates the dynamic range in stellar ages covered by our photometry.\label{fig:cmd_iso}}
\end{figure}

\subsection{Time Evolution of Stellar Structural Morphology}\label{s:ACFevol}

The sample of blue young stellar population of NGC\,6503 includes stars of 
various ages, and therefore represents different star formation events within the 
recent star formation history of the galaxy. This is demonstrated in Fig.\,\ref{fig:cmd_iso},
where the CMD of the blue stars is shown with indicative stellar evolutionary models for solar metallicity from
the Padova grid of models \citep[Chen et al., in prep.; see also][]{marigo08, girardi10, bressan12}.  
From the oldest isochrone it can be derived that the stellar age limit covered by our blue photometric catalog 
is \gsim\,100\,Myr (red line in Fig.\,\ref{fig:cmd_iso}). 

The stellar age range covered in our sample allows us to address the question whether star formation is a 
purely clustered process, or stars are also frequently born in a distributed fashion over galactic scales by investigating 
the clustering behavior of young stars at sequential evolutionary stages. The ACF treatment described in the previous section 
for the characterization of the clustering behavior of the whole blue sample can be also applied to stars of different ages. The 
construction of the ACF for subsamples of different ages would thus provide an assessment of how the clustering behavior of 
young populations evolves with time across the whole galaxy, over the last $\sim$\,100\,Myr, as investigated for example in the 
Magellanic Clouds \citep[]{gieles08, bastian09lmc}.

The derivation of ages of blue bright stars from their photometry alone is a complicated process and implies 
significant uncertainties for the derived parameters, even with the use of multi-band photometric measurements. 
In order, thus, to retain most of the original information derived from 
our catalogs, and to avoid introducing model-dependent uncertainties, we group the sample of blue
stars in ranges defined by their magnitudes in the F336W filter. We use this selection as a proxy for 
dividing the stars in groups of different evolutionary stages. In CMDs as this of  Fig.\,\ref{fig:cmd_iso}, 
both the turn-off (TO) and the He burning, starting immediately after the TO, correspond for every isochrone 
to a specific F336W magnitude which for older stars is systematically fainter. As a consequence, following  
the analysis by, e.g., \cite{bastian09lmc}, we group 
the stars within continuously fainter magnitude bins in order to roughly divide them into age bins. It should 
be noted though, that while the brightest magnitude range is populated only by the youngest stars, fainter 
ranges should also include  stars younger than the corresponding age limit. We assess, however, that the fraction
of younger contaminants is small enough not to significantly affect the canonical age of each magnitude bin\footnote{For 
a constant star formation rate, a 100 Myr old blue population and a disk of $\sim$\,10 Gyr, the young part 
in the fainter magnitude bins does not exeed $\sim$\,1\% of the old part.}.


\subsubsection{Two-Point Correlation Functions}\label{s:ACFevolsub}

\begin{figure}
\centerline{\includegraphics[clip=true,width=0.475\textwidth]{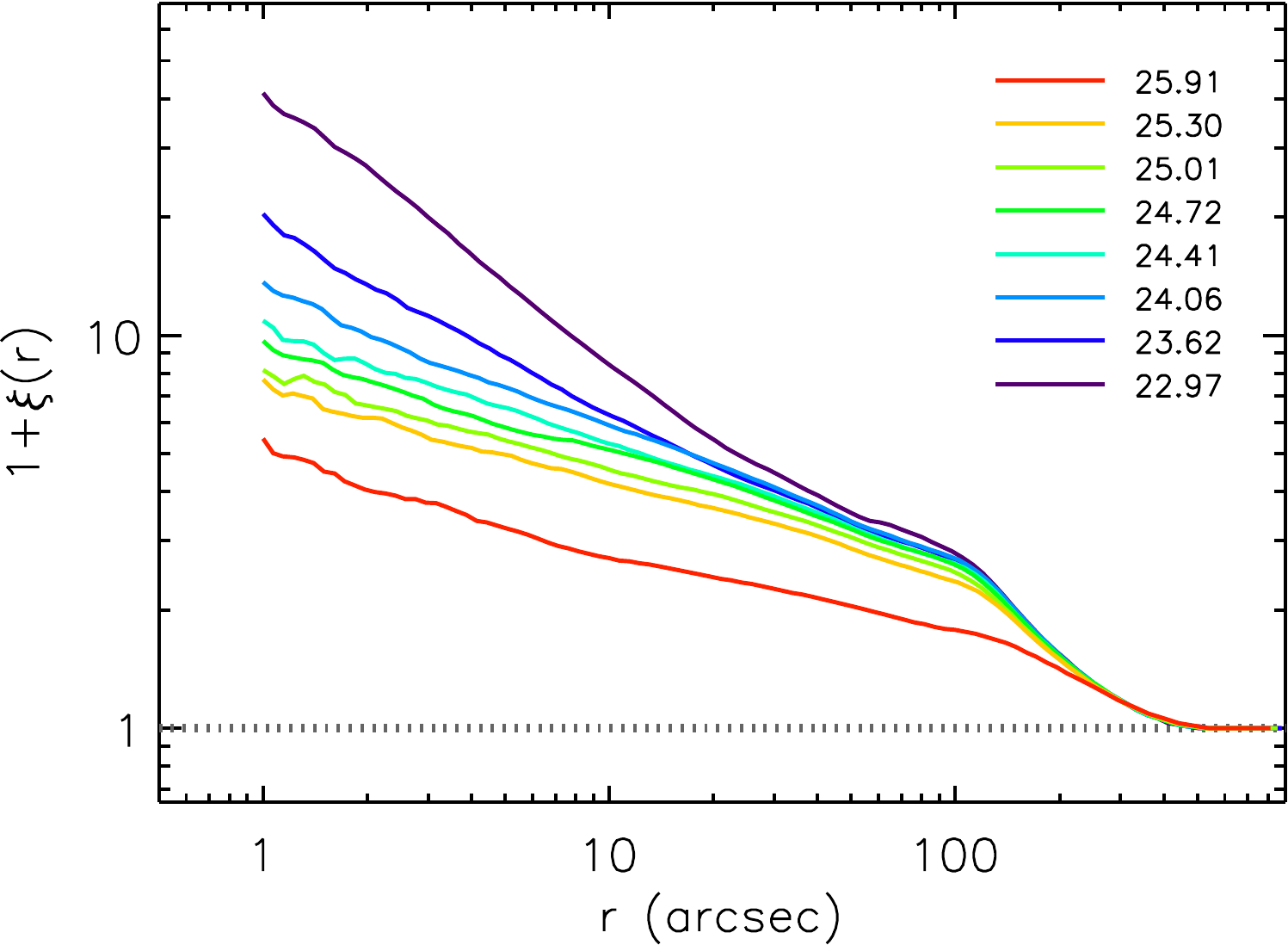}}
\caption{The two-point correlation (autocorrelation) function constructed for eight selected equally-numbered  
sub-groups of our sample of blue stars, corresponding to different magnitude intervals. In the plot legend  
the limit of each magnitude bin in the F336W filter is given next to the corresponding color line. 
\label{fig:aACF_magbin}}
\end{figure}

The blue stellar catalog is divided into eight magnitude bins, all containing equal numbers of $\simeq$\,1,600 stars.
Dividing the catalog into ranges of equal numbers satisfies equivalent statistical significance among all sub-samples. 
The limiting magnitudes of each sub-sample are shown in the first column of Table\,\ref{t:magbin}. The corresponding 
limiting age and stellar mass for each sub-sample, defined by the TO of the corresponding isochrone, are given in the 
second and third columns of the table respectively. We construct the ACF for each of the blue  sub-samples. The ACFs
are shown in Fig.\,\ref{fig:aACF_magbin} and the corresponding ACF exponent $\eta$ is given in Col. 4 of 
Table\,\ref{t:magbin}. Since some of the ACFs show broken power-law shapes, while others do not, we estimate 
$\eta$ up to the scale-length where all ACFs remain unchanged (for scales $\simeq$\,20\arcsec). From this analysis 
we find that indeed the brighter (and younger) stars in our sample are more clustered than the fainter ones. This is
demonstrated by both their systematically higher ACF values and steeper ACF slopes, which correspond to 
smaller $D_2$ and, thus, more clumpy distributions. 

The analysis described above also helps us designate the time-scale, where
a significant change in the clustering behavior of stars occurs. Specifically, $\eta$ for the three brightest 
magnitude intervals is quite different from one range to the other and from the remaining faint magnitude 
ranges. Specifically, $\eta$ becomes systematically shallower for stars with indicative ages between \lsim\,30\,Myr 
and $\sim$\,50\,or\,60\,Myr with values between $\sim$\,$-$0.7 and  $\sim$\,$-$0.3. However, from
this age on  $\eta$ remains almost unchanged with the value of $\sim$\,$-$0.2 up to $\sim$\,110\,Myr.
This result implies that while structure in the morphology of stellar clustering survives for the whole considered 
age range of $\lsim$\,100\,Myr, it changes significant its behavior within the first $\sim$\,60\,Myr. It is also
interesting to note that according to the ACF slopes found here, the clustering behavior of stars at the age 
limits $\sim$\,50\,-\,60\,Myr define the behavior of the whole sample the ACF of which has similar slope 
($\eta \sim$\,$-$0.3; Sect.\,\ref{s:ACF}). { The fitting uncertainty of the ACF 
exponent, $\eta$, is typically very small, ranging from less than 0.01 for stars in the brighter sub-sample
up to $\sim$\,0.03 for those in the fainter sub-sample. The measured errors are also given in Col. 4 of Table\,\ref{t:magbin}.}

\subsubsection{Pair and Minimum Spanning Tree Separations}\label{s:sepmst}

According to the results above, since younger stars show a more clumpy clustering behavior than the older,
they should be also typically confined in  smaller length-scales.  We confirm this result with the 
application of two measurements, the probability distribution function (PDF) of pair separations of 
all stars in each sub-sample, and the corresponding {\em minimum spanning tree} of these stars.   

We calculate the PDF of pair separations for all stars in each sample, 
as the sum of the probability functions of the stars, determined by the 
number of pair separations that fall in given separation bins around each 
star \citep[e.g.,][]{cw04}: 
\begin{eqnarray}\label{func:seppdf}
p(r_j) & = & \sum_{i=1}^{N} \frac{2\,N_{ij}}{N\,(N-1)\,dr} \,.
\end{eqnarray} 
In this function $N$ is the total number of stars,  $N_{ij}$ is the 
number of pair separations that fall in the separation bin centered 
on $r_j$ around star $i$, and $dr$ is the width of each separation bin. 
The probability that the projected separation between two randomly 
chosen stars is in the interval $(r, r+dr)$ is given by $p(r)dr$.  
The constructed pair separations PDFs of young stars show
different maxima for different magnitude ranges. Specifically,
the PDF becomes systematically more truncated toward smaller 
separations and its peak occurs at smaller length-scales  for
stars in younger sub-samples. The median of each derived PDF is
given in seconds of arc and kpc in Cols. 5 and 6 of Table\,\ref{t:magbin} 
respectively. While the RMS dispersion in each PDF is large enough to 
cover the median of its neighboring PDFs, these values show a clear trend 
of the typical pair separation of stars toward smaller lengths for higher stellar 
brightness and age. 

\begin{table}
\centering
\caption{Measures of the clustering behavior of young stars in eight distinct magnitude ranges, corresponding roughly to
different stellar age intervals between $\sim$\,4 and 110\,Myr. Magnitudes, ages, and masses listed in the first three columns 
correspond to the upper limits of each interval. The results of three different measurements are
given: The ACF slope $\eta$ (Col. 4); The median of all pair separations among stars in every sub-sample (Cols. 5
and 6); The median MST edge-length among stars in each sub-sample (Cols. 7 and 8). All measures indicate that the 
typical length-scale of stellar clustering depends on the brightness (age) range of the stars, with the brighter (younger) 
being concentrated in systematically smaller scales.}
\label{t:magbin}
\setlength\tabcolsep{0.1truecm}
\begin{tabular}{crccrccc}
\hline
{$m_{336}$} & 
{Age} &
{Mass} &
{$\eta$} &
\multicolumn{2}{c}{pair sep. median} &
\multicolumn{2}{c}{MST median} \\
{limit} & 
{(Myr)} &
{(M{\solar)}}&
{} &
{(arcsec)} &
{(kpc)} &
{(arcsec)} &
{(pc)}\\
\hline 
22.97	&	32	&	8.8	&	$-$0.693$\pm$0.004&	77.3	&	1.99	&	1.41	&	36.4	\\
23.62	&	40	&	7.7	&	$-$0.470$\pm$0.007&	80.2	&	2.06	&	1.85	&	47.6	\\
24.06	&	50	&	6.9	&	$-$0.321$\pm$0.009&	80.0	&	2.06	&	1.89	&	48.7	\\
24.41	&	63	&	6.2	&	$-$0.289$\pm$0.011&	82.4	&	2.12	&	2.04	&	52.4	\\
24.72	&	71	&	5.9	&	$-$0.232$\pm$0.012&	81.9	&	2.11	&	2.19	&	56.4	\\
25.01	&	89	&	5.3	&	$-$0.233$\pm$0.015&	84.5	&	2.17	&	2.33	&	60.0	\\
25.30	&	100	&	5.1	&	$-$0.225$\pm$0.016&	87.9	&	2.26	&	2.33	&	59.8	\\
25.91	&	112	&	4.9	&	$-$0.204$\pm$0.031&	104.9&	2.70	&	2.96	&	76.0	\\
\hline
\end{tabular}
\end{table}

The {\em minimum spanning tree} (MST), a construct from graph theory, is 
defined as the unique set of straight lines, called ``edges'', connecting a given 
set of points without closed loops, such that the sum of the edge lengths is 
the minimum \citep[e.g.,][]{kruskal56, prim57}. This method has been broadly 
used, along with other techniques, in clustering analysis of stellar samples 
\citep[see, e.g.][for a review]{schmeja11}. 
Here, we construct the PDF of the edge-lengths among stars in every sub-sample, 
as we did for their pair separations, as defined in Eq.\,(\ref{func:seppdf}). In this 
measurement instead of the pair separations among all stars, we consider the 
separations (edge-lengths) of each star from its nearest neighbors, as determined 
by the MST. The median of the MST edge-lengths for each stellar sample is given
in seconds of arc and parsecs in Cols. 7 and 8, respectively, of Table\,\ref{t:magbin}.
Again, there is a clear trend of younger stars being systematically clumped in smaller
scales as defined by the MST. 

\begin{figure}
\centerline{\includegraphics[clip=true,width=0.475\textwidth]{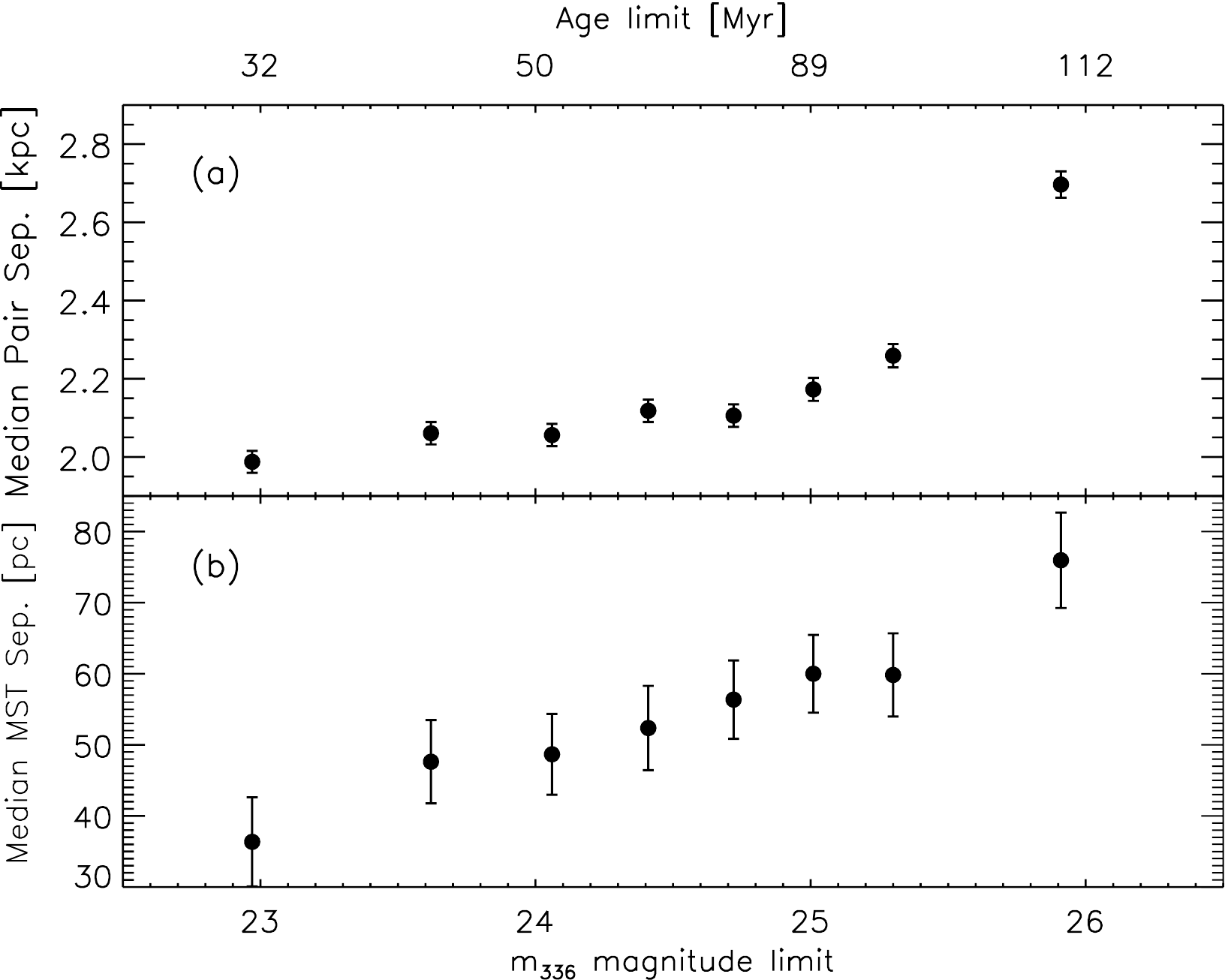}}
\caption{Relation between the medians of (a) the pair stellar separations and (b) the MST edge-lengths 
and the limiting magnitude (or age) for stars in eight selected brightness ranges (corresponding to 
different age intervals). These plots demonstrate that there is a dependence of the clustering 
length-scales of stars to their age. 
\label{f:agescaletrend}}
\end{figure}

The similarity of the scaling relations over stellar age 
found with both methods is demonstrated in Fig.\,\ref{f:agescaletrend}, where the 
corresponding medians are plotted in respect to the limiting magnitude and the equivalent  
age limit of each stellar sub-sample. { Typical uncertainties for the medians, corresponding to 
a few\,\%,  are also shown in Fig.\,\ref{f:agescaletrend}. They are calculated from the values corresponding to the data positions around
the median within the Poisson dispersion  of its position. }

While both measurements show the same trend of scale with age, the difference in the 
results between the pair separations and MST
edge-lengths, which reflects the difference of almost two orders of magnitude between 
the corresponding medians, lies on the fact that while the PDF of the pair separations 
is constructed from the whole extend of separations up to the highest covered length-scales,
that of the MST edge-length considers only the shortest edges between each star and its neighbors.
Under these circumstances, the medians given in Table\,\ref{t:magbin} for the MST edge-lengths 
represent the most likely typical scale of clustering for each stellar sample, while 
those for the pair separations correspond to the complete extent where stars 
in each magnitude bin are located over the whole observed field of view of the galaxy. The pair separations medians provide, 
thus, the ``inter-structure'' scaling length of stars at different ages, beyond the ``typical'' stellar clusterings, 
determined by the MST.

%



\section{Discussion}\label{s:disc}

The multi-scale sample of young stellar structures detected in NGC\,6503 suggests that 
star formation over a period of $\sim$\,100\,Myr takes place within the typical scale of few hundred parsecs (Sect.\,\ref{s:methclus}) 
in structures, which are themselves hierarchical as shown from their dendrogram (Sects.\,\ref{s:structtree}). 
The self-similar distribution of all young stars across the galaxy, demonstrated by their ACF, implies 
that hierarchical structure in the stellar distribution occurs not only inside but also beyond the borders 
of these structures. Therefore, one can conclude that all star-forming structures are connected to each other
in a hierarchical fashion through their coexistence in the star-forming ring of the galaxy (identified in Sect.\,\ref{s:ssdp}). 

Hierarchical structural morphology in stellar ensembles has been found by \cite{2elmegreen14} in 
the UV images of 12 LEGUS galaxies (NGC\,6503 not included in the sample) over a length-scale 
range of $\sim$\,1\,-\,200\,pc. The power-law length-scale correlations in the size and flux distribution 
functions of nested star-forming regions suggest that hierarchically structured star-forming 
regions with sizes of few hundred parsecs represent common unit structures. This is in line with 
our findings concerning the typical length-scale of the identified young stellar structures in NGC\,6503. 
Hierarchical structure was identified by \cite{2elmegreen14} only inside and not among the 
different star-forming patches, in both large spiral galaxies and low surface brightness dwarfs, as well as 
in starburst dwarfs or \hii\ galaxies, which are dominated by one or two large star-forming regions. 
In the case of NGC\,6503 self-similar stellar distributions are also detected among the star-forming structures,
suggesting that they are all connected to each other in a hierarchical fashion within the dominant  
star-forming ring of the galaxy.

\begin{figure}
\centerline{\includegraphics[clip=true,width=0.475\textwidth]{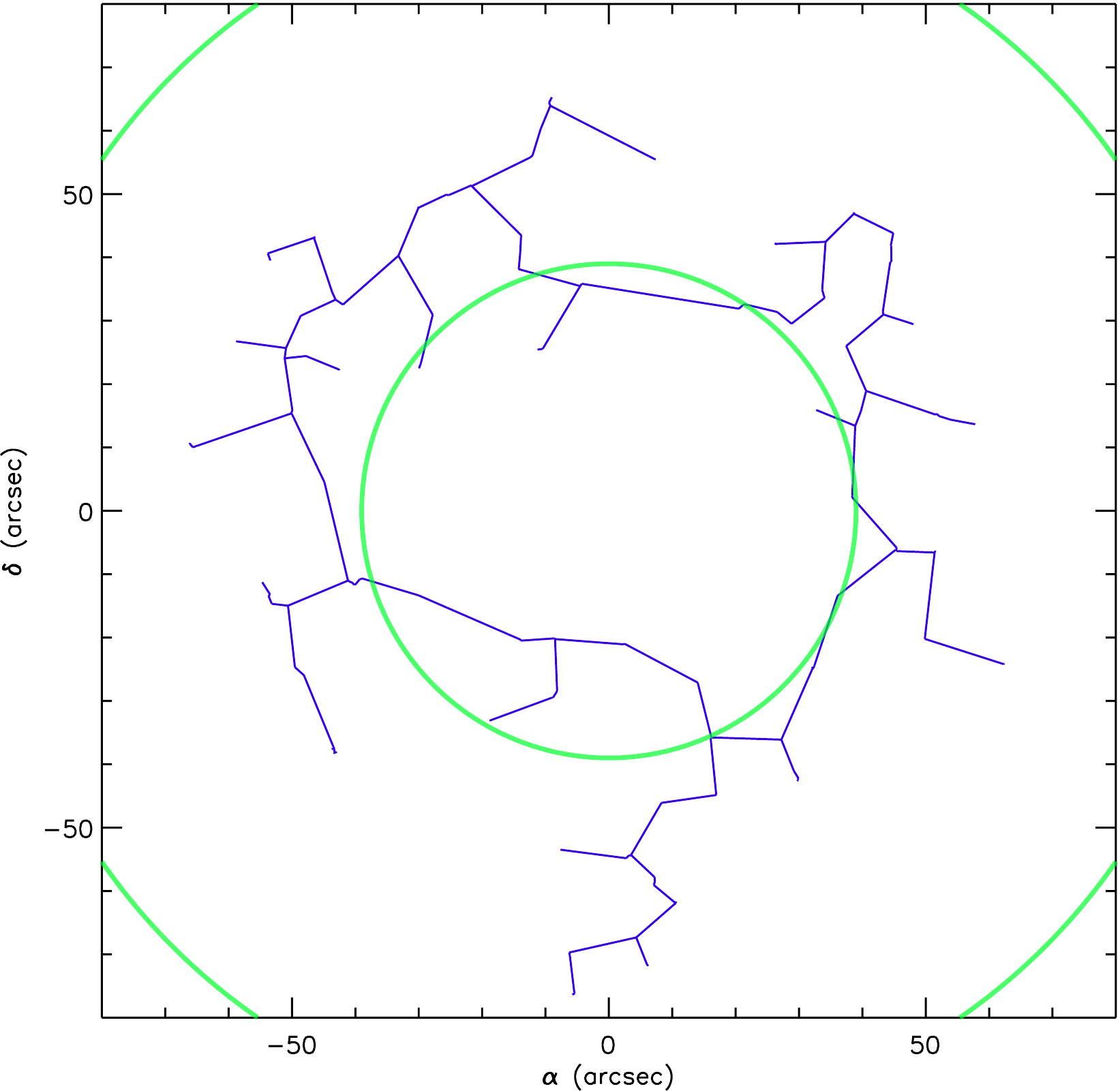}}
\caption{Minimum Spanning Tree connecting the positions of young stellar structures 
detected at significance levels higher than 3$\sigma$ (blue line). The MST connects each object to its 
nearest neighbor, drawing thus the positional alignment of these structures in respect to
the star-forming ring borders shown with the green circles.
\label{f:strctmst}}
\end{figure}

The observed hierarchical structure of star-forming complexes in NGC\,6503 and other galaxies is consistent 
with the model where star formation is regulated by turbulence, via, e.g., {\em turbulent fragmentation}, i.e., 
gas compressions that form successively smaller clouds inside larger ones \citep{vazquez-semadeni-09}. 
Such processes form a similar hierarchy of young stars, with a likely secondary correlation for star age, 
making larger regions older in proportion to the turbulent crossing time \citep{efremov-elmegreen-98, 
2de-la-fuente-marcos09}. Our findings for stars in different evolutionary stages confirm this time-length 
correlation, with younger stars being confined in smaller length-scale than the older. The hierarchy in 
stellar structures has an upper limit in size beyond which separate regions form independently \citep{2elmegreen14}. 
In the case of NGC\,6503 this limit is set by the size of the star-forming ring of the galaxy. This is consistent with 
the observation that the two-point correlation for stars decreases as a power law with increasing scale up to about 
2.5\,kpc.

The peak observed in the  young stellar surface density profile of Fig.\,\ref{f:radprof} corresponds to the star-forming ring of the 
galaxy, and thus its wings define the radial limits of the ring. These limits suggest an inner radius 
of $\sim$\,1\,kpc, and an outer radius of $\sim$2.5\,kpc. We derive from the stellar rotational curve of NGC\,6503 
\citep{bottema89} the projected asymptotic circular velocity for these radii to be of the order of $\sim$\,110\,km\,s$^{-1}$
at the { outer ring radius} and $\sim$\,80\,km\,s$^{-1}$ at the inner, corresponding to rotational velocities $\Omega_{\rm out} 
\simeq 1.4 \times 10^{-15}$\,rad\,s$^{-1}$ and $\Omega_{\rm inn} \simeq 2.6 \times 10^{-15}$\,rad\,s$^{-1}$ respectively.
The rotational velocity difference between the edges of the ring of about $10^{-15}$\,rad\,s$^{-1}$ is comparable to
the pattern speed of Milky Way's spiral arms \citep{bissantz03}. It most probably introduces a shear that should  
influence strongly the star formation process across it. 

This shear within $\sim$\,100\,Myr, which is the time covered by our observed young stellar sample, 
would have stretched a single compact object inwards, providing that this object is dense enough
to survive. The surface stellar density maps of Figs.\,\ref{fig:kde_maps} and \ref{fig:blue_kde_map} 
suggest an alignment of the detected young stellar structures across the ring of NGC\,6503. This trend 
is visualized by the MST of all structures found at significance levels $\geq$\,3$\sigma$, shown in 
Fig.\,\ref{f:strctmst}. In this figure the connecting edge-lengths show an alignment that both follows 
and traverses the ring closer to its inner part. Considering that the ring has performed several rotations during the last 
100\,Myr\footnote{A fit to the stellar rotational curve of NGC\,6503 \citep{bottema89}, based on the fitting formula by \cite{courteau97}, 
yields a projected asymptotic circular velocity of 108\,$\pm$\,4\,km\,s$^{-1}$, and a projected 
length-scale of 0.95\,$\pm$\,0.04\,kpc \citep{puglielli10}, suggesting that the inner part of the ring has performed more than 3 full rotations
during the last 100\,Myr.}, it stands to reason that shear functions more as a supportive rather than destructive factor to star formation.
The observed clustering of young stars in NGC\,6503 can be thus interpreted as being induced by turbulence, the driving source for which 
is probably gravitational instabilities induced by shear, with larger structures becoming spiral-like (flocculent) because of their longer dynamical 
time-scales in comparison to the shear time \citep[e.g.,][]{elmegreen11}.

\section{Summary}\label{s:sum}

We present a detailed clustering analysis of the young blue stellar population identified 
with LEGUS across the star-forming ring galaxy NGC\,6503. 
We construct stellar surface density maps and apply a contour-based analysis technique to identify the 
stellar complexes population of the galaxy. 
We identify 244 distinct structures at various stellar density (significance) levels. 
We organize these structures into 82 separate groups, according to their association with a 
single low-density parental structure. These groups are  arranged into eight families of structures, 
corresponding to eight super-structures,  determined by the 1$\sigma$  isopleths. 
Three of these super-complexes contain 95\% of the structures. 
The hierarchical classification of structures into groups and families, according to their membership to larger 
and sparser stellar constellations, is illustrated by their {\em dendrogram}, or {\em structure tree}.

We determine structural parameters, i.e., size, stellar density and total brightness, for each structure. The 
sizes of the detected systems average around $\sim$\,130\,($\pm$\,40)\,pc, a length-scale comparable to but still 
larger than $\sim$\,80\,pc, the scale which is discussed as a characteristic galactic scale for star formation 
\citep[see, e.g.,][and references therein]{gouliermis11}. About 60\% of the observed young stellar population is 
found to belong to one of the 1$\sigma$ super-structures,  suggesting that $\sim$\,40\% of the young blue stars 
is distributed in an ``unclustered" fashion inside the star-forming ring. The first statistically 
significant sample of stellar concentrations, detected at the 3$\sigma$ density level, accounts for only 34\% of the total 
observed young stellar content, increasing the remaining fraction of ``un-clustered'' young stars to 66\%.

On average, the size (and its dispersion), the total UV brightness, and the fraction of included young stars show a 
dependence on the density level where the corresponding structures are detected. The fraction of UV light included in
the structures over the total observed stellar UV emission shows a steeper dependence on density than that of the stellar 
fraction. This difference, which becomes more important for higher densities, suggests that compact stellar structures, identified 
at higher density levels, encompass on average the UV-brightest stars in the galaxy. 

 We identify a power-law mass-size relation, determined by the 
correlation between radius and number of stars. The exponent of this 
relation  for all structures is less than 2 as expected for a fractal distribution of  stars \citep[e.g.,][]{elmefalga96}. 
Our findings suggest a dependence of the power-law index of the mass-size relation to the detection threshold, 
in agreement with previous studies in the Large Magellanic Cloud \citep{gouliermis03} and M\,33 \citep{bastian07}. 
A correlation between the volume stellar density and the size of the structures was also found. This relation for the systems detected at 5 and 6$\sigma$ 
levels resembles Larson's
third relation \citep{larson81} and becomes steeper for higher-density structures, possibly reflecting their steeper density profiles. 
Since the volume density scales inversely with size, the identified correlation implies a dependence of
the parameters determination for the structures on the detection criteria, in a similar manner as for GMCs. 

The radial surface density profile of the young stars in NGC\,6503 shows a very strong central peak and 
drops rapidly within a radial distance of $\sim$\,350\,pc, in agreement with the line-of-site velocity dispersion 
profile, which shows a sharp decrease within 300\,pc \citep[][]{bottema89, bottema97}.
The stellar density profile rises again and 
reaches its peak at a distance of $\sim$\,1.9\,kpc, after which drops again exponentially. This profile, demonstrating
a circumnuclear deficiency in young stars, agrees with both an inner-truncated disk and a ring being formed by a bar \citep{puglielli10}. 
The confinement, though, of all young stellar structures in a ring-shaped alignment 
favors the ring-formation scenario, as earlier suggested \citep[][]{freeland10}.

We characterize the hierarchy in the global clustering behavior of young stars across NGC\,6503 with the autocorrelation 
function (ACF). We find that the ACF of these stars shows the typical features of a self-similar stellar distribution with a two-dimensional 
fractal dimension $D_{2} = 1.7$. The observed hierarchy in young stellar clustering extends monotonically 
across the complete measurable dynamic range in length-scales of two orders of magnitude ($\sim$\,20\,-\,2\,500\,pc).
The self-similar distribution of young stars across NGC\,6503 is consistent with the hierarchical morphology of the 
star-forming complexes of the galaxy, and the mass-size relation of 
the structures, which fits to the expectations for fractal clusterings. 

 We investigate the clustering behavior of young stars at
different evolutionary stages with the ACF of stars in different magnitude intervals. We find that younger (brighter) stars are more clustered
than the older. The ACF exponents of the younger stars are systematically higher, corresponding to smaller fractal dimensions and more clumpy 
distributions, than those for the older stars. This analysis shows that the time-scale, where
a significant change from a more clustered to a more distributed assembling of stars takes place, is about 60\,Myr. A similar trend
 was found for the Small and Large Magellanic Clouds, with a time-scale for sub-structure evolution towards 
a uniform distribution of $\sim$\, 75 and 175 Myr respectively \citep{gieles08, bastian09lmc}. Larger scale occupancy by 
stars of decreasing luminosity is previously discussed also for the spiral NGC\,1313 and dwarf irregular IC\,2574  \citep{pellerin07, pellerin12}.
These results are confirmed with the pair separations and Minimum Spanning Tree edge-lengths probability distributions of the stars 
in the various magnitude (age) ranges. 

Based on kinematic arguments we assess that hierarchy in the young stellar clustering in NGC\,6503 is probably induced by turbulence, driven 
by shear in the ring of the galaxy. With this mechanism large stellar structures become {\em flocculent} because their dynamical 
time-scales exceed the shear time.

\section*{Acknowledgments}
D.A.G. kindly acknowledges financial support by the German Research Foundation (DFG) through grant GO\,1659/3-2. 
This research has made use of the NASA/IPAC Extragalactic Database (NED), which is operated by the Jet Propulsion Laboratory, 
California Institute of Technology, under contract with the National Aeronautics and Space Administration.
This research has made use of the SIMBAD database, operated at CDS, Strasbourg, France. 
Based on observations made with the NASA/ESA {\sl Hubble Space Telescope}, obtained from the data archive at the 
Space Telescope Science Institute (STScI). STScI is operated by the Association of Universities for Research in Astronomy, Inc.\ under 
NASA contract NAS 5-26555. These observations are associated with program GO-13364. Support for Program 13364 was provided by NASA 
through grants from STScI. This research made use of the TOPCAT\footnote{TOPCAT is available at the permalink \href{http://www.starlink.ac.uk/topcat/}{http://www.starlink.ac.uk/topcat/ }} 
application \citep{topcat2005}, and NASAs Astrophysics Data System (ADS) bibliographic services\footnote{Accessible at \href{http://adswww.harvard.edu/}{http://adswww.harvard.edu/} 
and \href{http://cdsads.u-strasbg.fr/}{http://cdsads.u-strasbg.fr/}}.


\end{document}